\documentclass[11pt,a4paper]{article}
\pdfoutput=1
\usepackage{jheppub}
\usepackage{rotating}
\usepackage{array}
\usepackage{amsmath}
\usepackage[normalem]{ulem}
\usepackage{slashed}
\usepackage{booktabs}
\usepackage[pdftex,table,dvipsnames]{xcolor}
\usepackage{units}
\usepackage{multirow}
\usepackage{url}
\usepackage{parskip}
\usepackage[all]{nowidow}
\usepackage{placeins}
\usepackage{bm}

\usepackage{xspace}
\usepackage{subfig}
\usepackage{color}

\usepackage{dsfont}
\usepackage{soul}
\usepackage{hyperref}

\allowdisplaybreaks

\newcommand{\nc}{\newcommand}

\nc{\beq}{\begin{equation}}  
\nc{\eeq}{\end{equation}}  
\nc{\beqa}{\begin{eqnarray}}  
\nc{\eeqa}{\end{eqnarray}}  
\nc{\bit}{\begin{itemize}}  
\nc{\eit}{\end{itemize}}

\preprint{Nikhef 2024-018}

\title{Darkonia at Colliders}

\author[1]{Yang Bai,}

\author[2,3]{Susanne Westhoff\,}

\affiliation[1]{Department of Physics, University of Wisconsin-Madison, Madison, WI 53706, USA}

\affiliation[2]{Institute for Mathematics, Astrophysics and Particle Physics, Radboud University, 6500 GL Nij\-megen, The Netherlands}

\emailAdd{yangbai@physics.wisc.edu}
\emailAdd{susanne.westhoff@ru.nl}

\abstract{Dark matter may form bound states in a dark sector with an attractive force between two dark matter particles. Searches for dark matter at colliders can differ dramatically from routine searches if bound states, dubbed darkonia, are produced and decay into visible Standard-Model particles. In this work, we use three representative models with scalar, pseudo-scalar, and vector force carriers to map out the darkonium signatures at both high-energy and low-energy colliders. Some of the bound states can be stable due to generalized parity and charge-conjugation symmetries, while others decay into light dark-force carriers, which subsequently can decay at a displaced vertex. New signatures with a mix of missing energy and multiple di-lepton or di-jet vertices reconstructing intermediate darkonium resonances are within reach at the LHC and Belle II.
}

\begin{document}

\maketitle

\flushbottom
\newpage
\section{Introduction}
\label{SEC:intro}
There are numerous reasons to anticipate new interactions of dark matter particles. Most notably, explaining the observed dark matter abundance today typically requires dark matter interactions with particles of the Standard Model (SM) that govern their evolution in the early universe~\cite{Gondolo:1990dk}. On the other hand, inconsistencies in the interpretation of small-scale galaxy structures can be addressed if dark matter particles interact among themselves, see Ref.~\cite{Adhikari:2022sbh} for a review. Long-range interactions could be mediated by a massless force carrier, similar to the electromagnetic force between electrons and protons in atoms. Short-range interactions are usually mediated by a massive force carrier, akin to the one-pion exchanged Yukawa force between two nucleons in the Standard Model. If the force is attractive, dark matter particles can form a spectrum of dark bound states, dubbed darkonia in analogy with the quarkonia in the Standard Model. 

The existence of darkonia has significant effects on the dark matter phenomenology. Bound-state formation (in addition to the Sommerfeld effect) can enhance the dark matter annihilation cross section~\cite{vonHarling:2014kha}, with dramatic consequences on thermal freeze-out in the early universe and indirect dark matter detection today~\cite{Hisano:2004ds,Cirelli:2007xd,Arkani-Hamed:2008hhe,Binder:2023ckj}. In direct detection experiments, bound states of a dark matter particle and a nucleus can mediate resonant scattering, leading to an enhanced modulation signal~\cite{Pospelov:2008qx,Bai:2009cd}. In this work, we focus on the impact of bound-state formation on the third main method of dark matter searches: production and decay at colliders. Unlike conventional dark matter searches for invisible particles through missing energy, the production of dark bound states at colliders typically results in a characteristic pattern of visible particles in the final state. Darkonium decays generate new signatures, many of which are not covered by current searches.

In this work, we map out the landscape of darkonium signatures at colliders. Focusing on feebly interacting dark matter with GeV-to-TeV masses, we calculate the production and decay of darkonia made of fermion constituents for three representative models. In this way, we cover a wide range of possible darkonium signatures at proton-proton and electron-positron colliders, with direct applications at the Large Hadron Collider (LHC) and the flavor experiment Belle II.

Our analysis generalizes the results of a long history of studying bound states of hidden-sector particles. In supersymmetric theories, nearly mass-degenerate superpartners can form bound states whose decays produce characteristic signatures at high-energy colliders like the LHC~\cite{Nappi:1981ft,Barger:1988sp,Martin:2008sv}. Dark sectors with strongly interacting particles through a confining force similar to QCD can produce ``dark showers'' with signatures like emerging jets~\cite{Schwaller:2015gea} or semi-visible jets~\cite{Cohen:2015toa}. More recently, bound-state formation from a non-confining feeble force has been shown to predict interesting collider signatures~\cite{Shepherd:2009sa,An:2015pva,Tsai:2015ugz,Krovi:2018fdr}. The models we consider share features with this last class of bound states with feeble interactions.

The properties of darkonia at colliders depend on the mass of the dark matter constituent $m_\chi$, the mass of the dark force carrier $m_d$, the interaction strength $\alpha_d$ and potential symmetries in the dark sector. In the case of small couplings, $\alpha_d \ll 4\pi$, one can analyze the properties of dark bound states analogously to positronium, replacing the Coulomb force with a Yukawa force. Depending on the principal and partial-wave quantum numbers of the states, bound states exist for $m_d \lesssim \alpha_d \,m_\chi/2$, which favors a relatively light force carrier compared to the dark matter particle. In this work, we concentrate on dark matter mo\-dels with a weak short-range force. We postulate generalized parity and charge-conjugation symmetries for the interactions in the dark sector and find that, unlike for positronium, the lightest darkonium states which are odd under these symmetries are stable. An interesting addition would be to study collider signatures with $\alpha_d$ transitioning from weak to strong, connecting to the QCD-like case at $\alpha_d \sim 4\pi$.

The collider signatures of dark bound states depend on how the dark sector interacts with the Standard Model. We focus on three so-called portal interactions, which couple the dark sector through a single mediator particle in a gauge-invariant way~\cite{Pospelov:2007mp}: the scalar portal, where a new scalar boson couples to the Higgs field; the similar pseudo-scalar portal; and the vector portal, where a new gauge boson mixes with the hypercharge field. For the last model, bound-state production at Belle II has been studied before~\cite{An:2015pva}. We explore novel signatures at high-energy colliders like the LHC, including Drell-Yan production of vector darkonia via their mixing with photons or the $Z$ boson, as well as scalar bound-state production through Higgs mixing or Higgs boson decay. We also point out interesting new signatures at electron-positron colliders like Belle II, where bound states can be produced from $B$ meson decays.  

To calculate the darkonium production and decay rates at colliders, we introduce the Bound State Effective Field Theory (BSEFT), which describes the effective interactions of bound states with SM particles and among themselves. One advantage of using the BSEFT to study dark bound states is that it generalizes to the case where the binding force is independent of the mediator force. In this case, the interactions in the BSEFT parametrize our ignorance of the underlying dark matter models. For the minimal portal scenarios, where the mediator to the Standard Model is identical to the binding force, we provide the matching conditions between the underlying model couplings and the couplings in the BSEFT. We believe that the general BSEFT could also be efficient in studying the phenomenology of dark bound states beyond colliders.   

Our paper is organized as follows. In Sec.~\ref{SEC:properties}, we first describe the properties of dark bound states, including their binding energy and wave functions, in the three representative dark matter models. The BSEFT is introduced in Sec.~\ref{SEC:bs-eft}, where we spell out the Lagrangians for all three models and calculate the relevant matching conditions and formulas for bound-state decays. The collider phenomenology of the darkonia is presented in Sec.~\ref{SEC:collider}, where we analyze the main signatures in each model. We conclude this paper in Sec.~\ref{SEC:conclusions} and give a brief outlook to darkonium searches at future colliders. In Appendix~\ref{APPEND:wave-function}, we present detailed derivations of the relations between the BSEFT couplings and the bound-state wave functions.

\section{Properties of dark bound states}
\label{SEC:properties}
We start by defining the dark matter particles and their fundamental interactions, which lead to bound-state formation. In Sec.~\ref{SEC:models}, we introduce three dark sectors with fermion dark matter, distinguished by the force carrier that couples the dark matter particles among each other and to the SM sector. Subsequently, in Sec.~\ref{SEC:bound-state-mass}, we calculate the spectrum of dark fermion bound states and the corresponding wave functions. While the calculations in this section are model-specific, the methods we use can be applied more generally to bound-state formation in dark sectors with a perturbative, attractive scalar or vector force.

\subsection{Dark sector models}
\label{SEC:models}
We assume that the constituent particles of the bound states are Dirac fermions $\chi$. They are neutral under the SM gauge interactions and couple to massive mediator particles of a short-range force. We will consider both spin-zero (pseudo-)scalars or spin-one gauge bosons as force-carrier particles. Furthermore, we will also introduce some minimal interactions between the dark sector and the SM sector. These interactions are important for the properties of the dark bound states and potential signatures at colliders. 

The dark fermion is charged under its own discrete or continuous symmetries such that it is a stable particle. In this work, we ensure dark matter stability by introducing a $Z_2$ symmetry in the dark sector, under which $\chi \rightarrow - \chi$. The SM particles and the dark force carriers are $Z_2$-even. Furthermore, we assume that spacetime parity $P$ and charge conjugation symmetry $C$ are unbroken in the dark sector, although the weak interactions in the Standard Model break them explicitly.

Throughout this work, we restrict ourselves to 
 interactions that are invariant under two symmetries in the dark sector, dark parity $P_d$ and dark charge conjugation $C_d$. They are defined by
 \begin{align}\label{eq:dark-C-and-P}
P_d\,\chi_L P_d & = - \chi_R~,\quad \ P_d\,\chi_R P_d = - \chi_L \\\nonumber
C_d\,\chi\,C_d & = i \gamma^2 \, \chi^\ast~.
 \end{align}
 These symmetries are combinations of space-time parity $P$ or charge conjugation $C$ and the discrete $Z_2$ matter symmetry, similar to the $G$-parity in QCD~\cite{Lee:1956sw}. Bound states transform under $P_d$ and $C_d$ as under $P$ and $C$. If the mediator particle is $P$- and $C$-even, the lightest bound state which is odd under $P_d$ or $C_d$ can be stable and a potential dark matter candidate, similar to what happens in composite dark matter models with a dark $G$-parity~\cite{Bai:2010qg,Antipin:2015xia}. This class of darkonia are qualitatively different from the positronia or quarkonia in the Standard Model, which are all unstable. Notice that $P$- and $C$-violating interactions in the Standard Model do not break $P_d$ and $C_d$ and do not affect the stability of the bound states. 
 
Wherever relevant, we comment on possible extensions in $P_d$- and $C_d$-violating models.

\paragraph{Dark sector interactions}
In the case of scalar and pseudo-scalar force carriers, the masses and parity- and charge-conjugation-conserving interactions within the dark sector are described by
\beqa
\mathcal{L}_{\rm dark} &\supset& - m_\chi \overline{\chi} \chi - \frac{1}{2}m_d^2\,S^2 - g_d \,S\,\overline{\chi} \chi  \qquad \ \ \ \qquad (\mbox{F}_{\rm S}\;\mbox{model}) 
\label{eq:FS-basic-lag}
\\
\mathcal{L}_{\rm dark} &\supset& - m_\chi \overline{\chi} \chi - \frac{1}{2}m_d^2\,P^2 - g_{d 5} \,P\,\overline{\chi}\, i\gamma_5\, \chi  ~,  \qquad (\mbox{F}_{\rm P}\;\mbox{model})
\label{eq:FP-basic-lag}
\eeqa
where $m_\chi$ is the dark matter particle mass; $m_d$ is the mass of the force carriers $S$ and $P$; and $g_d$ and $g_{d5}$ are two types of Yukawa couplings. Note that $S$ is $P_d$-even and $P$ is $P_d$-odd, while both mediators are $C_d$-even. For later convenience, we denote the two models as $\mbox{F}_{\rm S}$ and $\mbox{F}_{\rm P}$, respectively.

In the case of vector gauge boson force carriers, the Lagrangian contains 
\beqa\label{eq:fv-lagrangian}
\mathcal{L}_{\rm dark} \supset - m_\chi \overline{\chi} \chi - \frac{1}{2}m_d^2\,A^\mu_d A_{d\,\mu}  - g_d\,A^\mu_d \, \overline{\chi} \gamma_\mu \chi ~.  \qquad (\mbox{F}_{\rm V}\;\mbox{model})
\eeqa
The massive dark gauge boson $A^\mu_d$ has a vector coupling $g_d$ to the dark fermions (see Ref.~\cite{Krovi:2018fdr} for a discussion including the axial-vector coupling). We denote this model as the $\mbox{F}_{\rm V}$ model. 

\paragraph{Interactions with the Standard Model}
We assume that the dark sector interacts with the Standard Model exclusively through the force carriers $S$, $P$ or $A_d$. Such a scenario corresponds to the widely studied (pseudo-)scalar and vector portals~\cite{Pospelov:2007mp}.

In the $\mbox{F}_{\rm S}$ and $\mbox{F}_{\rm P}$ models, the following renormalizable $P_d$- and $C_d$-conserving interactions between the dark sector and the SM sector are possible 
\beqa\label{eq:scalar-portal}
\mathcal{L}_{\rm portal} &=& - \mu_S \, S\,(H^\dagger H - \frac{v^2}{2}) - \lambda_S\,S^2 \, H^\dagger H ~ + \mu^2 H^\dagger H -\lambda (H^\dagger H)^2  \quad (\mbox{F}_{\rm S}\;\mbox{model})
\label{eq:portal-fermion-scalar} \\
\mathcal{L}_{\rm portal} &=& 
- \lambda_P\,P^2 \, H^\dagger H ~,  \qquad \qquad\qquad \qquad\quad\ \ \qquad \qquad \qquad (\mbox{F}_{\rm P}\;\mbox{model})
\label{eq:portal-fermion-pseudo-scalar}
\eeqa
where $H$ is the electroweak Higgs doublet in the Standard Model with vacuum expectation value (vev) $v = 246$~GeV. For later purposes, we have included the Higgs potential in~\eqref{eq:portal-fermion-scalar}.

In \eqref{eq:portal-fermion-scalar}, we have made a few assumptions and simplifications to the most general possible Lagrangian. We assume no vev for the scalar field $S$ and subtract the Higgs vev contribution from the first term, which would introduce a scalar vev through the linear term in $S$. The so-obtained interactions between the scalar $S$ and the Higgs boson $h$ are effectively the same as in a model where the dark scalar field obtains a vev, see e.g.~\cite{Freitas:2015hsa}.  We neglect a possible linear term in $S$ and self-interaction terms $S^3$ and $S^4$.\footnote{In Sec.~\ref{sec:fs-model}, we comment on the impact of self-interactions on bound-state decays.} These assumptions are purely phenomenologically motivated. They do not impact the dominant darkonium production channels at colliders, while allowing us to study the darkonium phenomenology in terms of a few parameters.

Following our symmetry assumptions, we have also neglected the possible parity-violating interaction $P\,H^\dagger H$ in~\eqref{eq:portal-fermion-pseudo-scalar}.

In the $\mbox{F}_{\rm V}$ model, the dark sector can interact with the SM sector through kinetic mixing with the hypercharge field,
\beqa
\mathcal{L}_{\rm portal} = - \frac{1}{2}\frac{\epsilon}{c_w} B_{\mu\nu}F_d^{\mu\nu}\ . ~\qquad (\mbox{F}_{\rm V}\;\mbox{model})
\label{eq:portal-fermion-vector}
\eeqa
Here, $F_d^{\mu\nu} = \partial^\mu A_d^{\nu} - \partial^\nu A_d^\mu $ is the dark gauge field tensor; $B_{\mu\nu}$ is the hypercharge gauge field tensor; and $c_w \equiv \cos{\theta_w}$ with $\theta_w$ the weak mixing angle. Through kinetic mixing, the interaction states $B$, $W^{3}$ and $A_d$ are related to the mass eigenstates for the photon ($A$), $Z$ boson ($Z$), and dark photon ($A'_d$), as 
\begin{align}
\begin{pmatrix} B \\ W^3 \\ A_d \end{pmatrix} & = 
    \begin{pmatrix} 1 & 0 & -\eta \\
    0 & 1 & 0 \\
    0 & 0 & \eta/\hat{\epsilon} \end{pmatrix}
    \begin{pmatrix} c_w & - c_\xi s_w & s_\xi s_w \\
    s_w & c_\xi c_w & - s_\xi c_w \\
    0 & s_\xi & c_\xi    \end{pmatrix}
\begin{pmatrix} A \\ Z \\ A'_d \end{pmatrix},
\end{align}
with the kinetic mixing parameters $\hat{\epsilon} = \epsilon/c_w$ and $\eta = \hat{\epsilon}/\sqrt{1-\hat{\epsilon}^2}$. The mixing angle $\xi$ is defined by
\begin{align}
    \tan(2\,\xi) = \frac{2\,\eta\, s_w}{1-(\eta\,s_w)^2 - \delta}\ ,\qquad \delta = (1-\hat{\epsilon}^2)\frac{m_d^2}{m_Z^2}\ ,\qquad m_Z = \frac{g\,v}{2\,c_w} \ .
\end{align}
 In the limit of small kinetic mixing $\eta$ and mass ratio $\delta$, the mixing angle is approximated by $\xi \approx \epsilon\tan\theta_w$. The masses of the dark photon and $Z$ boson receive only second-order corrections in $\eta$ and $\delta$; we will neglect them. The photon remains massless. In this limit, the interaction eigenstates are related to the mass eigenstates by
\begin{align}\label{eq:field-rotations-fv-model}
B & = c_w A - s_w Z - \epsilon\, c_w A'_d \\\nonumber
W^3 & = s_w A + c_w Z - \epsilon\, s_w A'_d \\\nonumber
A_d & = A'_d + \epsilon\, t_w Z \ .
\end{align}
After properly normalizing the gauge field, the dark photon couples to SM fermions as
\begin{align}
    \mathcal{L} \supset \epsilon\,e\,A'^\mu_d \sum_f Q_f \bar{f} \gamma_\mu f \,,
\end{align}
with $Q_f$ denoting the electric charge of the SM fermion $f$.

Note that in the three representative models $\mbox{F}_{\rm S}$, $\mbox{F}_{\rm P}$, $\mbox{F}_{\rm V}$, the dark force mediator also acts as the portal particle to the SM sector. We call this case the \emph{minimal scenario}. In extended dark sectors, called \emph{general scenarios}, the dark force mediator responsible for bound-state formation can be different from the portal particle connecting the dark sector to the Standard Model. In particular, non-Abelian dark forces can lead to a different darkonium phenomenology.

\subsection{Bound-state masses and wave functions}
\label{SEC:bound-state-mass}
For the darkonium phenomenology at colliders, we consider bound states that are made of one dark fermion and one dark antifermion. Stable bound states can also be formed from two dark fermions or two dark antifermions if the binding force is scalar. Such bound states could have non-trivial cosmological consequences, but are more challenging to be produced at colliders. The reason is that, to conserve the dark fermion number, a total of four particles must be produced — two dark fermions and two dark antifermions. Since four-particle production at colliders is phase-space suppressed, fermion-fermion or antifermion-antifermion bound states are less likely to be produced.

Using the notation of $^{2S+1}L_J$ and $J^{PC}$, we consider the following lowest bound states for the three models:  
\begin{align}
  & \bm{\eta_d}:\ ^1S_0,\  0^{-+}\,,\qquad\bm{\Upsilon_d}:\ ^3S_1,\  1^{--} \,, \qquad
\bm{h_d}:\ ^3P_0,\  0^{++} 
\qquad (\mbox{F}_{\rm S}) \\\nonumber
&   \bm{\eta_d}:\ ^1S_0,\  0^{-+} \ \ \hspace*{7.5cm} (\mbox{F}_{\rm P}) \\\nonumber
&\bm{\eta_d}:\ ^1S_0,\  0^{-+}\,,\qquad\bm{\Upsilon_d}:\ ^3S_1,\  1^{--}\ .
\hspace*{4cm}\!  (\mbox{F}_{\rm V})
\end{align}
We include all bound states that can be singly produced at colliders. For example, in the $\mbox{F}_{\rm S}$ model, the $0^{++}$ bound state $h_d$\footnote{Here, the notation is different from the charmonium and bottomonium systems, where the $0^{++}$ state is denoted as $\chi_{c0}$ or $\chi_{b0}$.} can mix with the force carrier $S$ and directly couple to the SM sector. 

In the $\mbox{F}_{\rm S}$ and $\mbox{F}_{\rm V}$ models, the Yukawa-like static potential between the dark fermion and antifermion can be
attractive and universally defined as
\beqa
V(r) = - \frac{\alpha_d}{r} \, e^{- m_d\,r} ~,
\label{eq:potential}
\eeqa
with the dark coupling strength $\alpha_d \equiv g_d^2/(4\pi)$.

In the $\mbox{F}_{\rm P}$ model, the pseudo-scalar force is spin-dependent, with a static potential proportional to the combination of dark fermion spins $\vec{s}_{\bar{\chi}} \cdot \vec{s}_\chi$. As such a potential is only attractive for a total spin $S = 0$ state, we will only consider the pseudo-scalar state $\eta_d$ in the $\mbox{F}_{\rm P}$ model. The calculation of the static potential is similar to the di-nucleon force from one-pion exchange, except without the isospin symmetry. So, we define $\alpha_d \equiv g_{d5}^2/(4\pi) \times m^2_d /(4 m^2_\chi)$ for the $\mbox{F}_{\rm P}$ model~\cite{Epelbaum:2008ga}, where the parameter dependence originates from the derivative coupling nature of the pion as a Goldstone boson and the Goldberger-Treiman relation.

\begin{figure}[t!]
\centering
    \includegraphics[width=0.48\textwidth]{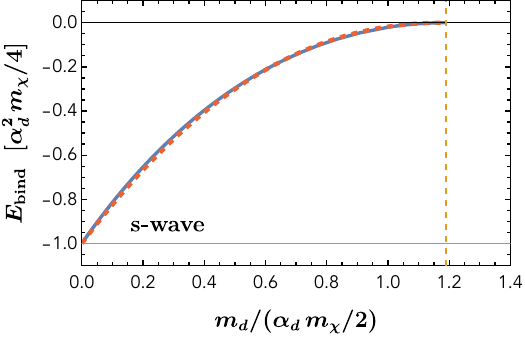} \hspace{3mm}
    \includegraphics[width=0.48\textwidth]{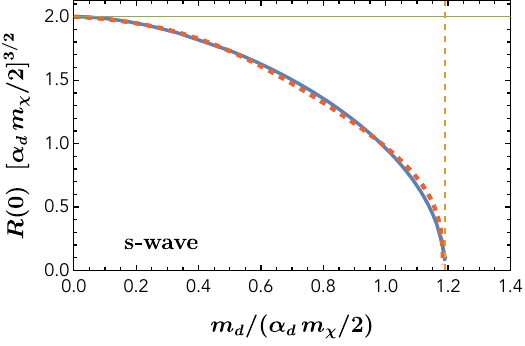}
    \caption{{\it Left panel:} The binding energy of darkonium as a function of the force-carrier mass $m_d$ in units of the Bohr momentum $\alpha_d\,m_\chi/2$ for the lightest s-wave bound state. The numerically fitted function defined in \eqref{eq:binding-energy-fit} is shown as a red dashed curve. {\it Right panel:} The radial wave function at origin, in the same presentation as in the left panel. The numerically fitted function can be found in \eqref{eq:wave-origin-fit}. S-wave bound states exist only when $m_d/(\alpha_d\,m_\chi /2) < 1.19$, as indicated by the vertical dashed orange line. The horizontal green lines correspond to a massless force-carrier, similar to the positronium case.
    \label{fig:bindingenergy-s-wave}}
\end{figure}

For all models, we solve the Schr\"odinger equation using the static potential in \eqref{eq:potential} to obtain the bound-state masses and wave functions (see Refs.~\cite{Ennes:mathematica,Edwards:2017ndv} for a similar calculation).

\paragraph{S-wave bound states} only exist for a mediator mass $m_d$ below around the Bohr momentum of the system~\cite{PhysRev.125.1131},
\begin{align}\label{eq:bs-condition-s}
    \frac{m_d}{\alpha_d\,m_\chi/2} < 1.19~.
\end{align}
In the left panel of Fig.~\ref{fig:bindingenergy-s-wave}, we show our numerical solution of the Schr\"odinger equation for the binding energy $E_{\rm bind}$ as a function of the ratio of $m_d$ over the Bohr momentum. One can see that the binding energy decreases towards zero as the ratio approaches the upper limit of the bound-state condition~\eqref{eq:bs-condition-s}. In the other limit with a massless mediator, $m_d \rightarrow 0$, one recovers the positronium case with $E_{\rm bind} = - \alpha_d^2 \,m_\chi /4$.

For later convenience, we fit the numerically calculated results for the binding energy as
\beqa
E_{\rm bind}({\rm s}-{\rm wave}) = - \frac{\alpha^2_d \, m_\chi}{4}\, S_E\left(\frac{m_d}{\alpha_d \,m_\chi/2}\right) \quad \mbox{with} \quad S_E(x) =  - \left(1- \dfrac{x}{1.19}\right)^{2.2}, \label{eq:binding-energy-fit}
\eeqa
shown as a red dashed curve in the left panel of Fig.~\ref{fig:bindingenergy-s-wave} for comparison. 

In the right panel of Fig.~\ref{fig:bindingenergy-s-wave}, we show the radial wave function at the origin, $R(0)$, as a function of $m_d/(\alpha_d m_\chi/2)$. The radial wave function obeys the normalization condition $\int dr\, r^2 R^2(r) = 1$. The total wave function at the origin is $\psi(0) = R(0)/\sqrt{4\pi}$ for the s-wave state. When the mediator mass reaches the upper limit, $R(0)$ becomes zero, which indicates a flat and unnormalized state. In the opposite limit with $m_d \rightarrow 0$, the wave function at origin matches the positronium case with $\psi(0)=(\alpha_d\,m_\chi/2)^{3/2}/\sqrt{\pi}$.

The result for the wave function is fitted as
\beqa
\psi(0) &=& \frac{R(0)}{\sqrt{4\pi}} = \left(\frac{\alpha_d\,m_\chi}{2}\right)^{3/2}\, S_{\psi_0}\left(\frac{m_d}{\alpha_d \,m_\chi/2}\right) \nonumber \\
&&\qquad\qquad\qquad\qquad \mbox{with} \quad S_{\psi_0}(x) = \frac{1}{2}\,\left[ \arctan\big((x^{-1} - 0.845)^{1/3}\big) \right]^{1/4}~,\label{eq:wave-origin-fit} 
\eeqa
which is shown as a red dashed curve in the right panel of Fig.~\ref{fig:bindingenergy-s-wave}.

\begin{figure}[t!]
\centering
    \includegraphics[width=0.48\textwidth]{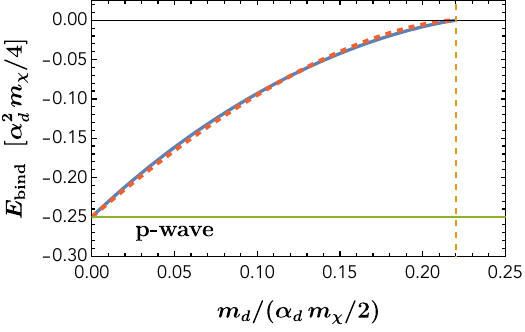} \hspace{3mm}
    \includegraphics[width=0.48\textwidth]{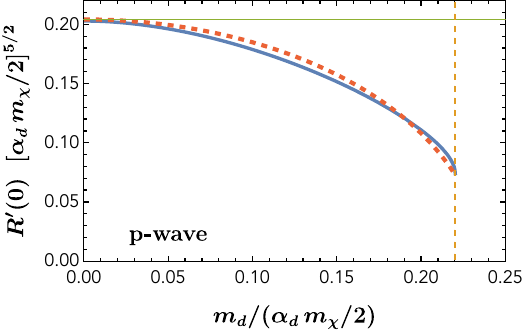}
    \caption{{\it Left panel}: The binding energy of a darkonium as a function of the force-carrier mass $m_d$ in units of the Bohr momentum $\alpha_d\,m_\chi/2$ for the lightest p-wave state. The numerically fitted function defined in \eqref{eq:binding-energy-fit-p} is shown as a red dashed curve. {\it Right panel}: The derivative of the radial wave function at the origin, in the same presentation as in the left panel. The numerically fitted function can be found in \eqref{eq:wave-origin-fit-p}. P-wave bound states exist only when $m_d/(\alpha_d\,m_\chi /2) < 0.22$, as indicated by the vertical dashed orange line. The horizontal green lines correspond to a massless force-carrier, similar to the positronium case.
        }
    \label{fig:bindingenergy-p-wave}
\end{figure}

\paragraph{P-wave bound states} For p-wave bound states, the condition to have a bound state is more stringent than in the s-wave case~\cite{PhysRevA.1.1577}
\begin{align}\label{eq:bs-condition-p}
    \frac{m_d}{\alpha_d\,m_\chi/2} < 0.22~.
\end{align}
In the left panel of Fig.~\ref{fig:bindingenergy-p-wave}, we show the binding energy as a function of $m_d/(\alpha_d\,m_\chi/2)$ for p-wave darkonium with the numerically fitted function as 
\beqa
E_{\rm bind}({\rm p}-{\rm wave}) = - \frac{\alpha^2_d \, m_\chi}{4}\, P_E\left(\frac{m_d}{\alpha_d \,m_\chi/2}\right) \quad \mbox{with} \;\; P_E(x) =  - \frac{1}{4}\,\left(1- \dfrac{x}{0.22}\right)^{1.6}.\label{eq:binding-energy-fit-p}
\eeqa
Again, the binding energy approaches zero as the mediator mass reaches the upper limit in \eqref{eq:bs-condition-p}. In the massless limit $m_d \rightarrow 0$, one recovers the positronium binding energy of $E_{\rm bind} = - \alpha^2 m_\chi/16$ for the $2p$ state.

Since the p-wave function at the origin is zero, we calculate the first derivative of the radial wave function at the origin, which will be relevant for the production and decay of p-wave darkonium. We show our numerical solution in the right panel of Fig.~\ref{fig:bindingenergy-p-wave}, together with the fitted function
\beqa\label{eq:fs-derivative}
R'(0) = \left(\frac{\alpha_d\,m_\chi}{2}\right)^{5/2}\, P_{R'_0}\left(\frac{m_d}{\alpha_d \,m_\chi/2}\right) \quad \mbox{with} \quad P_{R'_0}(x) = \frac{1}{\sqrt{24}}\,\left( 1- 18 x^2 \right)^{1/2} ~.\label{eq:wave-origin-fit-p} 
\eeqa
For a massless mediator with $m_d = 0$, one recovers the result for the Hydrogen wave-function derivative, $R'(0) = (\alpha_d\,m_\chi/2)^{5/2}/\sqrt{24}$ with $\alpha_d \to \alpha$ and $m_\chi \to m_e$.

For the s-wave bound states $\eta_d$ and $\Upsilon_d$, hyperfine splitting makes $\Upsilon_d$ heavier than $\eta_d$ based on the Yukawa potential. The mass splitting is suppressed by $\alpha_d^4$~\cite{Berko:1980gg,Eichten:1980mw} in the massless limit $m_d\to 0$  and even more suppressed for a nonzero mediator mass. This has consequences for the collider phenomenology, as we will discuss in Sec.~\ref{sec:fv-colliders}.

\subsection{Dark matter relic abundance}

Throughout this work, we will be agnostic about the relic abundance of both the dark matter constituents and the dark matter bound states. In this way, our predictions of darkonium production and decay can be generally applied to scenarios with or without a dark matter context. In this section we comment on freeze-out scenarios that could explain the dark matter relic abundance through early-universe dynamics.

In general, the dark matter abundance today can be a mixture of dark fermion constituents and (stable) bound states. In the ${\rm F}_{\rm S}$ and ${\rm F}_{\rm P}$ models, the lightest $P_d$- and $C_d$-odd states are stable and contribute to the total dark matter energy density.

The thermal freeze-out of the dark constituents follows the standard calculation~\cite{Gondolo:1990dk}, including potential effects of bound state formation~\cite{vonHarling:2014kha}. In the absence of bound state effects, the leading annihilation channel of a dark fermion-antifermion pair is into two force mediators. Using the ${\rm F}_{\rm S}$ model as an example, the annihilation rate for $m_\chi \gg m_S$ reads
\begin{align}
    \sigma(\chi \bar{\chi} \rightarrow S S)\,v = \frac{3 g_d^2}{128\pi m_\chi^2}\,v^2
\end{align}
and is $p$-wave suppressed. Here, $v \sim 0.3$\,c is the dark fermion speed at around the freeze-out temperature. To satisfy the total relic abundance, one has $3 g_d^2/(128\pi m_\chi^2) \approx 5.5$\,pb~\cite{Bai:2012qy} or $m_\chi/g_d \approx 735\,$GeV. The formation of bound states can significantly enhance the annihilation cross section and allow for much lighter dark matter candidates. Lighter dark matter is also favored in the scenario where the constituents only make up a fraction of the total dark matter abundance.


The abundance of stable bound states can be calculated in a similar way as for hydrogen formation. In analogy with the recombination process $e^- p \rightarrow H \gamma$, the relic bound state abundance is determined by processes like $\chi \bar{\chi} \rightarrow \eta_d  S$ in the ${\rm F}_{\rm S}$ model. Since the mediator $S$ is massive, the rate of bound state formation depends on the model parameters, e.g. the relation between binding energy and $S$-particle mass. The formation of fermion-fermion bound states could further complicate the total relic abundance calculation. Given the scope of this work, we do not explore the origin of the bound state abundance in detail.

Finally, the LHC phenomenology places greater emphasis on the portal coupling, which is not as crucial for bound state formation in the early universe. Therefore, one has the flexibility to reduce the coupling to comply with current LHC constraints, while still expecting detectable signals at the HL-LHC and future experiments.

\section{Effective theory for bound-state interactions}
\label{SEC:bs-eft}
The interactions of dark bound states can be described by an effective theory, which we call the Bound State Effective Field Theory (BSEFT).\footnote{See Ref.~\cite{Kilic:2008pm} for a similar approach with colored bound states.} Two relevant scales determine the effective interactions: the mass of the lightest dark bound state, $M$; and the relative velocity of the bound-state constituents, $v = |\vec{p}_\chi|/m_\chi$, defined in the rest frame of the bound state. The mass $M$ determines the cutoff scale of the BSEFT, $\Lambda_d \approx M$.

Conceptually, the description of bound-state formation at $v \ll 1$ in the BSEFT is similar to NRQCD~\cite{Bodwin:1994jh}. We restrict ourselves to contributions at the leading order in $v$. The scale $M$ separates low-energy and high-energy momenta in processes involving bound states. Decays of dark bound states only involve particle momenta up to $M/2$. The production of bound state can happen at momenta much larger than $M$, provided that the relative velocity $v$ of the constituents in sufficiently small. In this case, the BSEFT couplings implement the factorization and projection approach described in Ref.~\cite{Petrelli:1997ge}. Both decay and production of darkonia can therefore be calculated within the BSEFT.

When constructing the BSEFT Lagrangian, we do not make any assumptions about the force that binds the dark fermions into bound states. The structure of the BSEFT is thus valid regardless of the underlying model. The coefficients of the interactions, however, are model-dependent. For the minimal scenario, where the binding force carrier is also the mediator to the Standard Model, we calculate some of these coefficients using the bound-state wave functions from Sec.~\ref{SEC:bound-state-mass}.

\subsection{$\mbox{F}_{\rm S}$ model}\label{sec:fs-model}
In the $\mbox{F}_{\rm S}$ model, the CP-even bound state $h_d$ can mix with the force-mediator field $S$ that couples to the SM Higgs field. For operators with mass dimension 4 or lower, the BSEFT Lagrangian for darkonium couplings reads 
\begin{align}\label{eq:bseft-FS}
\mathcal{L}_{\rm BSEFT}^{(4)} & = \frac{1}{2}\partial_\mu \eta_d \partial^\mu \eta_d - \frac{m_{\eta_d}^2}{2}\eta_d^2 - \frac{1}{4} \Upsilon_d^{\mu\nu}\Upsilon_{d,\mu\nu}   + \frac{m_{\Upsilon_d}^2}{2}\Upsilon_d^\mu \Upsilon_{d,\mu} \nonumber \\
& \quad + \frac{1}{2}\partial_\mu S \partial^\mu S + \frac{1}{2}\partial_\mu h_d \partial^\mu h_d - \frac{m_S^2}{2}SS - g_d\,\mu_d^2\,h_d\,S - \frac{m_{h_d}^2}{2}\,h_d^2\\\nonumber
& \quad + \lambda_h\,S\,S\,h_d + \lambda_h'\,S\,h_d\,h_d
 + \lambda_{\eta_d}'\,S\,\eta_d\,\eta_d
 + \xi_{\eta_d}\,h_d\,\eta_d\,\eta_d \ .
\end{align}
Following our rationale from Sec.~\ref{SEC:models}, we have included all $P_d$- and $C_d$-conserving interactions with up to three particles. In particular, we have neglected the $P_d$- and/or $C_d$-violating couplings $SS \eta_d$, $S h_d\eta_d$, $S \Upsilon_d \eta_d$, $S h_d \Upsilon_d$ and $h_d h_d \eta_d$. Notice that because $\Upsilon_d$ is $C_d$-odd and $\eta_d$ is $P_d$-odd, all dark-sector interactions with one single $\Upsilon_d$ or one single $\eta_d$ violate $C_d$ or $P_d$ symmetry, respectively. Requesting these symmetries therefore implies that $\Upsilon_d$ and $\eta_d$ are both stable.

 Couplings with two $\Upsilon_d$ fields like $S \Upsilon_d^\mu \Upsilon_{d,\mu}$ and $h_d \Upsilon_d^\mu \Upsilon_{d,\mu}$ are in principle allowed by the symmetries of the underlying model. However, amplitudes based on these couplings violate partial-wave unitarity at energies $E \gg m_{\Upsilon_d}$. Unitarity can be restored by treating $\Upsilon_d$ as a fundamental particle that receives its mass from spontaneous symmetry breaking. Alternatively, the composite nature of $\Upsilon_d$ can be accounted for by introducing a form factor which regularizes the high-energy behavior of the BSEFT (see also the related discussion in Ref.~\cite{Freitas:2014jla}). Since both approaches lead to model-dependent results, we do not include these couplings in our analysis.

Couplings with more than three particles have been neglected in~\eqref{eq:bseft-FS}, as they are not relevant for the collider phenomenology described in Sec.~\ref{SEC:collider}. They can, however, play a role in radiation and scattering processes that involve two or three bound states. 

\paragraph{Determining BSEFT parameters} The $h_d-S$ mixing strength $g_d\,\mu_d^2$ in~\eqref{eq:bseft-FS} is related to the wave function properties of the bound state. In the minimal scenario where $S$ is the binding force and the mediator to the SM sector, the scalar mixing is proportional to the radial derivative of the p-wave function at the origin from the right panel of Fig.~\ref{fig:bindingenergy-p-wave}. We calculate the mixing strength by performing a non-relativistic expansion of the corresponding field operator (see Appendix~\ref{APPEND:wave-function} for the derivation)
\beqa
g_d\,\mu_d^2 = g_d \,\frac{\sqrt{2m_{h_d}}}{m_\chi}\, \frac{3}{\sqrt{2\pi}}\,R'_{h_d}(0)\ .
\eeqa

The interaction terms in the BSEFT Lagrangian can be obtained by calculating an appropriate amplitude in the underlying theory and matching it onto the corresponding amplitude in the BSEFT. To obtain the amplitude in the underlying theory, we factorize contributions above and below the BSEFT cutoff scale $\Lambda_d$ and use projection techniques developed for NRQCD~\cite{Petrelli:1997ge} to project onto the bound state with the desired quantum numbers. 

To determine the coupling $\lambda_h$ in~\eqref{eq:bseft-FS}, we calculate the elementary process $\chi\bar{\chi} \to SS$ and project the $\chi\bar{\chi}$ pair onto the bound state $h_d$. The decay rate for $h_d \to SS$ in the underlying theory is then given by
\begin{align}\label{eq:hd-to-s-s}
\Gamma(h_d\to SS) = 8\,\alpha_d^2\,\frac{|R_{h_d}'(0)|^2}{m_{h_d}^4}\bigg(1-\frac{4 m_S^2}{m_{h_d}^2}
\bigg)^{\frac{5}{2}}\bigg(1-\frac{2 m_S^2}{m_{h_d}^2}\bigg)^{-4}.
\end{align}
By matching the squared matrix element~\footnote{In the case of the p-wave bound state $h_d$, the comparison between the underlying theory and the BSEFT cannot be done at the amplitude level. The projection onto the $S = 0$ state $^3 P_1$ requires to square the amplitude. See Ref.~\cite{Petrelli:1997ge} for details.} for general four-momenta $p_1$, $p_2$ of the outgoing (on-shell or off-shell) scalars $S$ to the corresponding squared matrix element in the BSEFT, we identify the BSEFT coupling
\begin{align}\label{eq:lambdah}
    \lambda_h(p_1,p_2) = 8\sqrt{\pi}\alpha_d\frac{R_{h_d}'(0)}{m_{h_d}^{3/2}}\frac{[(2\,p_1\cdot p_2 - m_{h_d}^2 + p_1^2)\,p_1^2 + (p_1\cdot p_2)^2] + [p_1 \leftrightarrow p_2]}{(p_1\cdot p_2)^2}\ .
\end{align}
To ensure bound-state formation, the relative momentum of the dark-fermion constituents of $h_d$ must be small. This requirement translates into a condition on the three-momenta of the two scalars, which must fulfill $|\vec{p}_1 - \vec{p}_2| \ll m_{h_d}$.

A second contribution to $h_d \to SS$ can arise from $h_d - S$ mixing, $h_d \to S^\ast \to SS$, if a triple $S$ coupling is present. We do not include this model-dependent process here. 

\paragraph{Interactions with several bound states} The BSEFT couplings $\lambda_h'$, $\lambda_{\eta_d}'$ and $\xi_{\eta_d}$ of two or more bound states in~\eqref{eq:bseft-FS} are more difficult to determine than those of one single bound state. The production of two bound states via $h_d \to \eta_d\eta_d$, $S\to \eta_d\eta_d$ or $S\to h_d h_d$, cannot be accurately described in the BSEFT. Since the involved momentum lies above the cutoff scale $\Lambda_d \approx M$ of the BSEFT, relativistic corrections to bound-state production are expected to be numerically relevant. As a consequence, identifying the BSEFT coefficients by matching the amplitude in the underlying theory to the non-relativistic limit is phenomenologically not very useful. 

\begin{figure}[t!]
\centering
    \includegraphics[width=0.46\textwidth]{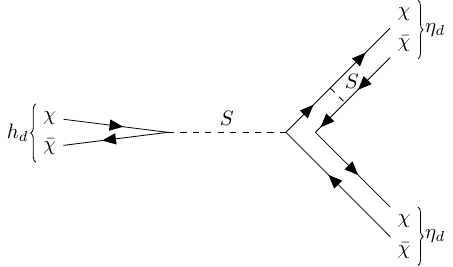} \hspace{3mm}
    \includegraphics[width=0.46\textwidth]{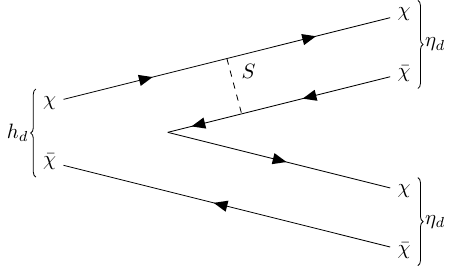} 
    \caption{Bound-state pair production via $h_d \to \eta_d \eta_d$ in the $\mbox{F}_{\rm S}$ model.}
    \label{fig:bs-pair-production}
\end{figure}

Despite these shortcomings, the BSEFT is still a valuable framework to estimate the relative rates of bound-state pair production.  Here we give an estimate of the decay rate for $h_d \to \eta_d\eta_d$. In the BSEFT, the decay width reads
\begin{align}\label{eq:hd-to-etad-etad}
    \Gamma(h_d \to \eta_d\,\eta_d) = \frac{|\xi_{\eta_d}|^2}{32\pi m_{h_d}}\bigg(1 - \frac{4 m_{\eta_d}^2}{m_{h_d}^2}\bigg)^{\frac{1}{2}}.
\end{align}
As said above, we cannot determine the coupling $\xi_{\eta_d}$ from amplitude matching. But we can determine the scaling of $\xi_{\eta_d}$ with the radial wave function $R(0)$ and the dark-sector coupling $g_d$ by qualitatively comparing with the decay amplitude in the underlying theory. Diagrammatically, there are two leading contributions, see Fig.~\ref{fig:bs-pair-production}.
 The first contribution scales as $\xi_{\eta_d} \propto g_d^4  R'_{h_d}(0) R_{\eta_d}^2(0) /m_\chi^{9/2} \propto \alpha_d^{15/2} m_\chi$, using $R'(0) \propto (\alpha_d m_\chi)^{5/2}$ and $R(0) \propto (\alpha_d m_\chi)^{3/2}$. The second contribution involves an $h_d\to \eta_d\eta_d$ transition form factor, which cannot be estimated from first principles, but could dominate the decay. We expect that the production of two bound states is suppressed compared to the decay into elementary force carriers, $h_d \to SS$, due to the binding effects.

The decays $S\to h_d h_d$ and $S\to \eta_d\eta_d$ are kinematically forbidden if $S$ is the mediator of the binding force and $\alpha_d$ is perturbative, see Sec.~\ref{SEC:collider}. We can still use the corresponding amplitudes to estimate the couplings $\lambda_h'$ and $\lambda_{\eta_d}'$ in~\eqref{eq:bseft-FS}. Unlike in the case of three bound states, the  production of two bound states can be described to first approximation in terms of wave functions~\cite{Keung:1983ac}. Using the $S$ decay in the left diagram in Fig.~\ref{fig:bs-pair-production}, we estimate
\begin{align}\label{eq:2-bs-couplings}
    \lambda_h' \propto \frac{g_d^3}{m_\chi^4} (R_{h_d}'(0))^2 \propto \alpha_d^{13/2}m_\chi\ ,\qquad 
    \lambda_{\eta_d}' \propto \frac{g_d^3}{m_\chi^2} (R_{\eta}(0))^2 \propto \alpha_d^{9/2}m_\chi\ .
\end{align}

\paragraph{BSEFT at mass dimension 5} At mass dimension 5 in the BSEFT, possible additional interactions are
\begin{align}\label{eq:bseft-FS-5}
\mathcal{L}_{\rm BSEFT}^{(5)} = \frac{C_{S\Upsilon}}{\Lambda_d} S\,\Upsilon_d^{\mu\nu}\Upsilon_{d,\mu\nu}
+ \frac{C_{h\Upsilon}}{\Lambda_d} h_d\, \Upsilon_d^{\mu\nu}\Upsilon_{d,\mu\nu} + \frac{C_{\eta\Upsilon}}{\Lambda_d}\eta_d\,\Upsilon_d^{\mu\nu}\widetilde{\Upsilon}_{d,\mu\nu} \ .
\end{align}
Again, we have imposed $P_d$ and $C_d$ symmetry, which explains the dual field-strength tensor $\widetilde{\Upsilon}_{d,\mu\nu}$ in the third term, rather than $\Upsilon_{d,\mu\nu}$.

The coupling $C_{h\Upsilon}$ induces the decay $h_d \to \Upsilon_d\Upsilon_d$, provided that $m_{h_d} > 2 m_{\Upsilon_d}$. We find the decay rate in the BSEFT to be
\begin{align}\label{eq:hd-to-upsd-upsd}
\Gamma(h_d \to \Upsilon_d\Upsilon_d) = \frac{1}{4\pi}\frac{|C_{h\Upsilon}|^2}{\Lambda_d^2}m_{h_d}^3\left(1 - \frac{4 m_{\Upsilon_d}^2}{m_{h_d}^2}\right)^{\frac{1}{2}}\left(1 - 4\frac{m_{\Upsilon_d}^2}{m_{h_d}^2} + 6 \frac{m_{\Upsilon_d}^4}{m_{h_d}^4}\right).
\end{align}
Due to binding effects, the effective coupling $C_{h\Upsilon}$ cannot be estimated from first principles, but should scale similarly to $\xi_{\eta_d}$ in $h_d \to \eta_d\eta_d$.
The couplings $C_{S\Upsilon}$ and $C_{\eta\Upsilon}$ scale as $\lambda_{\eta_d}'$ in~\eqref{eq:2-bs-couplings}. In the minimal scenario they are irrelevant for the collider phenomenology because the bound-state condition implies that decays $S \to \Upsilon_d \Upsilon_d$ and $\eta_d \to \Upsilon_d\Upsilon_d$ are kinematically forbidden.

\paragraph{Higgs portal mixing} Besides the $S-h_d$ mixing, a second source of scalar mixing arises from the Higgs portal~\eqref{eq:scalar-portal}. In the broken phase of the electroweak theory, the portal interactions read~\footnote{We do not consider a vacuum expectation value for $S$ and work in unitary gauge.}
\begin{align}\label{eq:higgs-portal-after-ewsb}
\mathcal{L}_{\rm portal} = - \mu^2 h^2 - \frac{\mu_S}{2} Sh\, (2v + h) - \frac{\lambda_S}{2}S^2h\, (2v + h)\ .
\end{align}
Including $h-S$ and $S-h_d$ mixing, the mass matrix for the three scalars $h$, $S$ and $h_d$ reads
\begin{align}
    \mathcal{L}_m = \frac{1}{2}\big(h\ S\ h_d\big)\mathcal{M}^2\begin{pmatrix} h \\ S \\ h_d \end{pmatrix},\qquad \mathcal{M}^2 = \begin{pmatrix} 2\mu^2 & \mu_S \, v & 0 \\
   \mu_S\, v & m_S^2 & g_d \,\mu_d^2 \\ 0 & g_d \,\mu_d^2 & m_{h_d}^2\end{pmatrix}\,.
\end{align}
For small $h-S$ and $S-h_d$ mixing, this matrix can be diagonalized by two separate rotations acting on the 1-2 and 2-3 sectors of $\mathcal{M}^2$, respectively. Introducing the small mixing angles $\theta_h$ and $\theta_{h_d}$, we can approximate the diagonalization matrix by
\begin{align}
   R \approx \begin{pmatrix} 1 & \theta_h & 0 \\ -\theta_h & 1 & \theta_{h_d} \\ 0 & -\theta_{h_d} & 1\end{pmatrix},\quad \theta_h = \frac{\mu_S \,v}{2\mu^2-m_S^2}\ ,\quad \theta_{h_d} = \frac{g_d \, \mu_d^2}{m_S^2-m_{h_d}^2}\ .
\end{align}
The mass eigenstates of the physical particles are given by
\begin{align}
\begin{pmatrix} \hat{h} \\ \hat{S} \\ \hat{h}_d \end{pmatrix} = R\begin{pmatrix} h \\ S \\ h_d \end{pmatrix}.
\end{align}
The (squared) mass eigenvalues of the physical scalars, $\hat{m}_h^2$, $\hat{m}_S^2$ and $\hat{m}_{h_d}^2$, are then obtained as
\begin{align}
    R\,\mathcal{M}^2 R^T = \text{diag}(\hat{m}_h^2,\hat{m}_S^2,\hat{m}_{h_d}^2) \,,
\end{align}
with the mass eigenvalues
\begin{align}
    \hat{m}_h^2 & = 2\mu^2 + (2 \mu^2 - m_S^2) \theta_h^2\\\nonumber
    \hat{m}_S^2 & = m_S^2 - (2 \mu^2 - m_S^2) \theta_h^2 + (m_S^2 - m_{h_d}^2)\theta_{h_d}^2\\\nonumber
    \hat{m}_{h_d}^2 & = m_{h_d}^2 - (m_S^2 - m_{h_d}^2)\theta_{h_d}^2 \ .
\end{align}
In terms of mass eigenstates, the mass and interaction terms in~\eqref{eq:bseft-FS} and~\eqref{eq:higgs-portal-after-ewsb} read
\begin{align}\label{eq:fs-model-interactions}
    \mathcal{L}_{\rm BSEFT}^{(4)} & = - \frac{\hat{m}_{\eta_d}^2}{2}\hat{\eta}_d\hat{\eta}_d + \frac{\hat{m}_{\Upsilon_d}^2}{2}\Upsilon_d^\mu \Upsilon_{d,\mu} - \frac{\hat{m}_S^2}{2}\hat{S}\hat{S} - \frac{\hat{m}_{h_d}^2}{2}\hat{h}_d\hat{h}_d - \frac{\hat{m}_h^2}{2}\hat{h}\hat{h}  \\\nonumber
& \quad + (\lambda_S \,v\,\theta_h - \lambda_h\, \theta_{h_d}) \hat{S}^3 + \lambda_h'\, \theta_{h_d}\, \hat{h}_d^3\\\nonumber
& \quad + (\lambda_h - 2 \lambda_h' \,\theta_{h_d})\hat{S}\,\hat{S}\,\hat{h}_d + (\lambda_h' + 2 \lambda_h\, \theta_{h_d})\hat{S}\,\hat{h}_d\,\hat{h}_d \\\nonumber
& \quad - (\lambda_S\,v + \mu_S \,\theta_h) \hat{S}\,\hat{S}\,\hat{h}  + \lambda_h' \,\theta_h \,\hat{h}_d\,\hat{h}_d\,\hat{h} + 2\left(\lambda_h\,  \theta_h  - \lambda_S\, v\,\theta_{h_d} \right) \hat{S}\,\hat{h}_d\,\hat{h} \\\nonumber
& \quad + (\lambda_{\eta_d}' - \xi_{\eta_d} \,\theta_{h_d}) \hat{S}\,\eta_d\,\eta_d + (\xi_{\eta_d} + \lambda_{\eta_d}'\,\theta_{h_d} ) \hat{h}_d \,\eta_d \,\eta_d + \lambda_{\eta_d}' \,\theta_h \,\hat{h} \,\eta_d \,\eta_d ~,
\end{align}
up to corrections of $\mathcal{O}(\theta_h^2,\theta_{h_d}^2,\theta_h\theta_{h_d})$. The stable states $\eta_d$ and $\Upsilon_d$ couple to the Standard Model only through the Higgs field and in pairs, see the last two lines of~\eqref{eq:fs-model-interactions}. The scalar bound state $\hat{h}_d$ inherits all couplings of the Higgs to SM particles, but doubly suppressed by scalar mixing $\theta_h\theta_{h_d}$. In particular, the coupling to fermions is described by
\begin{align}
\mathcal{L} = - \frac{m_f}{v/\sqrt{2}}\sin\theta_h\sin\theta_{h_d} \,\bar{f}f \hat{h}_d\ ,
\end{align}
where $m_f$ is the mass of fermion $f$.

In bound-state decays, SM couplings play essentially no role, since $\hat{h}_d$ decays preferentially into binding-force carriers for $\hat{m}_{h_d} > 2 \hat{m}_S$. Bound-state production, however, often relies on the coupling to fermions.

Finally, the dimension-5 interactions from~\eqref{eq:bseft-FS-5} in terms of mass eigenstates are given by
\begin{align}\label{eq:fs-interactions-dim5}
\mathcal{L}_{\rm BSEFT}^{(5)} & = \left(\frac{C_{S\Upsilon}}{\Lambda_d} - \frac{C_{h\Upsilon}}{\Lambda_d} \theta_{h_d}\right) \hat{S}\,  \Upsilon_d^{\mu\nu}\Upsilon_{d,\mu\nu} + \left(\frac{C_{h\Upsilon}}{\Lambda_d} + \frac{C_{S\Upsilon}}{\Lambda_d} \theta_{h_d}\right) \hat{h}_d\,\Upsilon_d^{\mu\nu}\Upsilon_{d,\mu\nu} \\\nonumber
& \quad + \frac{C_{S\Upsilon}}{\Lambda_d} \theta_h \hat{h}\Upsilon_d^{\mu\nu}\Upsilon_{d,\mu\nu}  + \frac{C_{\eta\Upsilon}}{\Lambda_d}\eta_d\,\Upsilon_d^{\mu\nu}\widetilde{\Upsilon}_{d,\mu\nu} ~.
\end{align}

In the bound-state decay rates~\eqref{eq:hd-to-s-s}, \eqref{eq:hd-to-etad-etad} and \eqref{eq:hd-to-upsd-upsd}, we have neglected scalar mixing in the couplings and mass terms. Including it would introduce small corrections of $\mathcal{O}(\theta_h^2,\theta_{h_d}^2)$.

\subsection{$\mbox{F}_{\rm P}$ model}
In the simpler $\mbox{F}_{\rm P}$ model, we only have one s-wave bound state, $\eta_d$. Using the relation of $\alpha_d \equiv g_{d5}^2/(4\pi) \times m^2_d /(4 m^2_\chi)$ and the bound-state condition in \eqref{eq:bs-condition-s}, one has a stringent bound on the ratio of the mediator mass over the dark fermion mass 
\beqa
\label{eq:FP-model-constraint}
\frac{m_d}{m_\chi} > \frac{84.5}{g^2_{d5}} ~.
\eeqa
For $m_d$ lighter than $m_\chi$, as required in the minimal scenario, the coupling $g_{d5}$ is required to be non-perturbative (similar to the one-pion exchanged force between two nucleons), which calls the calculability of this model into doubt. In the general scenario, the BSEFT is still a useful guide for collider searches.

Independent of the binding force, we define the BSEFT Lagrangian 
\beqa\label{eq:bseft-F_P}
\mathcal{L}_{\rm eff} &=& \frac{1}{2}\partial_\mu \eta_d \partial^\mu \eta_d - \frac{m_{\eta_d}^2}{2}\eta_d\eta_d + \frac{1}{2}\partial_\mu P \partial^\mu P  - \frac{1}{2}m_P^2 P^2 - \mu_d^2\,g_{d5}\,\eta_d\,P  ~. 
\eeqa
Here, the mass parameter $m_P^2$ contains both the bare mediator mass square $m_d^2$ from \eqref{eq:FP-basic-lag}, as well as the mass contribution from the coupling to the Higgs doublet after electroweak symmetry breaking from~\eqref{eq:portal-fermion-pseudo-scalar}. Following our symmetry assumptions in~\eqref{eq:dark-C-and-P}, we have neglected the possible parity-violating term $P P\eta_d$.

For bound states with $P$ as the force mediator, the coefficient $\mu_d^2\,g_{d5}$ is related to the s-wave function at the origin through (see Appendix~\ref{APPEND:wave-function})
\beqa g_{d5}\,\mu_d^2 = 2\, g_{d5}\, \sqrt{\frac{m_{\eta_d}}{\pi}}\,R_{\eta_d}(0) = 4\,g_{d5}\,\sqrt{m_{\eta_d}}\,\psi_{\eta_d}(0) ~,
\eeqa
using the wave function at the origin, $\psi_{\eta_d}(0)$, from~\eqref{eq:wave-origin-fit}. 

Rotating the two states $\eta_d$ and $P$ into mass eigenstates
\begin{align}
\begin{pmatrix} \hat{P} \\ \hat{\eta}_d \end{pmatrix} = \begin{pmatrix} \cos\theta & -\sin\theta \\ \sin\theta & \phantom{+}\cos\theta \end{pmatrix} \begin{pmatrix} P \\ \eta_d \end{pmatrix},
\end{align}
with the mixing angle $\theta = \frac{1}{2}\,\arctan{\left[2\,\mu_d^2\,g_{d5}/(m_{\eta_d}^2 - m_P^2)\right]}$, yields the (squared) mass eigenvalues 
\begin{align}
\hat{m}_P^2 & =  \frac{1}{2}\, \left[ m_P^2 + m_{\eta_d}^2 - \sqrt{(m_{\eta_d}^2 - m_P^2)^2 + 4 g_{d5}^2\,\mu_d^4} \right] \\\nonumber
\hat{m}_{\eta_d}^2 & =  \frac{1}{2}\, \left[ m_P^2 + m_{\eta_d}^2 + \sqrt{(m_{\eta_d}^2 - m_P^2)^2 + 4 g_{d5}^2\,\mu_d^4} \right] \,.
\end{align}

Using the portal interaction in \eqref{eq:portal-fermion-pseudo-scalar}, electroweak symmetry breaking induces couplings of the bound state and the mediator to the SM Higgs boson,
\beqa
\mathcal{L}_{\rm portal} \supset - \lambda_P \, (\cos{\theta}\,\hat{P} + \sin{\theta}\,\hat{\eta}_d )^2\,\left(v\,h + \frac{1}{2}h^2\right) ~.
\eeqa
For the case with $\hat{m}_{\eta_d} > \hat{m}_P$, $\hat{P}$ is the lightest parity-odd particle and stable. The heavier state $\hat{\eta}_d$, on the other hand, can decay into $\hat{P}$ plus an off-shell Higgs boson.

\subsection{$\mbox{F}_{\rm V}$ model}\label{sec:fv-model}
In the $\mbox{F}_{\rm V}$ model, the vector state $\Upsilon_d$ mixes with the dark photon. The BSEFT Lagrangian at mass dimension 4 reads
\begin{align}\label{eq:bseft-dark-photon}
\mathcal{L}_{\rm BSEFT}^{(4)} = & - \frac{1}{4} F_d^{\mu\nu}F_{d,\mu\nu} + \frac{\epsilon_d}{2}g_d F_d^{\mu\nu}\Upsilon_{d,\mu\nu} - \frac{1}{4} \Upsilon_d^{\mu\nu}\Upsilon_{d,\mu\nu}\\\nonumber
& + \frac{1}{2}\partial_\mu \eta_d \partial^\mu \eta_d - \frac{m_{\eta_d}^2}{2}\eta_d\eta_d + \frac{m_{\Upsilon_d}^2}{2}\Upsilon_d^\mu \Upsilon_{d,\mu} + \frac{m_d^2}{2} A_d^\mu A_{d,\mu} ~,
\end{align}
with the field-strength tensors
\begin{align}
F_d^{\mu\nu} = \partial^\mu A_d^\nu - \partial^\nu A_d^\mu\ ,\quad \Upsilon_d^{\mu\nu} = \partial^\mu \Upsilon_d^\nu - \partial^\nu \Upsilon_d^\mu\ ,\quad \widetilde{V}^{\mu\nu} = \frac{1}{2}\epsilon^{\mu\nu\rho\sigma} V_{\rho\sigma}\ .
\end{align}
Due to kinetic mixing, the interaction state of the dark photon is a linear combination of mass states, $A_d  = A'_d + \epsilon\, t_w Z$, see~\eqref{eq:field-rotations-fv-model}.

The mixing term $F_d^{\mu\nu}\Upsilon_{d,\mu\nu}$ in~\eqref{eq:bseft-dark-photon} can be removed by the field transformations
\begin{align}\label{eq:field-trafo}
A'_d{}^\mu \to A'_d{}^\mu + g_d\epsilon_d \Upsilon_d^\mu \ ,\qquad
Z^\mu \to Z^\mu + \epsilon t_w g_d\epsilon_d \Upsilon_d^\mu ~,
\end{align}
up to corrections of $\mathcal{O}(g_d^2\epsilon_d^2)$.
These transformations lead to a non-diagonal mass matrix, so that
\begin{align}
\mathcal{L}_m = \frac{1}{2}\begin{pmatrix} Z^\mu & \Upsilon_d^\mu & A'_d{}^\mu \end{pmatrix} \mathcal{M}^2 \begin{pmatrix} Z^\mu\\ \Upsilon_{d,\mu} \\ A'_{d,\mu} \end{pmatrix},\qquad \mathcal{M}^2 = \begin{pmatrix} m_Z^2 & \epsilon\, t_w \, g_d\, \epsilon_d\, m_Z^2 & 0 \\
\epsilon\, t_w\, g_d\, \epsilon_d\, m_Z^2 & m_{\Upsilon_d}^2 & g_d \,\epsilon_d\, m_d^2 \\
0 & g_d\, \epsilon_d\, m_d^2 & m_d^2\end{pmatrix}.
\end{align}
For small mixing, the mass matrix $\mathcal{M}^2$ is diagonalized by separate rotations in the 1-2 and 2-3 sectors. In terms of the small mixing angles $\theta_Z$ and $\theta_{\Upsilon}$, the diagonalization matrix is approximated by
\begin{align}\label{eq:fv-mass-mixing}
   R \approx \begin{pmatrix} 1 & \theta_Z & 0 \\ -\theta_Z & 1 & \theta_{\Upsilon} \\ 0 & - \theta_{\Upsilon} & 1\end{pmatrix},\quad \theta_Z = \epsilon \,t_w \,g_d\, \epsilon_d \frac{m_Z^2}{m_Z^2 - m_{\Upsilon_d}^2}\ ,\quad \theta_{\Upsilon} = g_d\,\epsilon_d\, \frac{m_d^2}{m_{\Upsilon_d}^2 - m_d^2} ~.
\end{align}
In this approximation, the mass eigenstates $\hat{Z}$, $\hat{\Upsilon}_d$, $\hat{A}'_d$ are related to the intermediate states $Z$, $\Upsilon_d$, $A'_d$ after kinetic mixing by
\begin{align}
\begin{pmatrix} \hat{Z} \\ \hat{\Upsilon}_d \\ \hat{A}_d  \end{pmatrix} = R\begin{pmatrix} 1 & \epsilon\,t_w \,g_d \,\epsilon_d & 0 \\
0 & 1 & 0 \\
0 & g_d \,\epsilon_d & 1 \\
\end{pmatrix}
\begin{pmatrix} Z \\ \Upsilon_d \\ A'_d \end{pmatrix}.
\end{align}
The eigenvalues of the physical bosons are
\begin{align}\label{eq:-fv-mass-ev}
    \hat{m}_Z^2 & = m_Z^2 + (m_Z^2 - m_{\Upsilon_d}^2) \theta_Z^2 \\\nonumber
    \hat{m}_{\Upsilon_d}^2 & = m_{\Upsilon_d}^2 - (m_Z^2 - m_{\Upsilon_d}^2) \theta_Z^2  + (m_{\Upsilon_d}^2 - m_d^2) \theta_{\Upsilon}^2 \\\nonumber
    \hat{m}_d^2 & = m_d^2 - (m_{\Upsilon_d}^2 - m_d^2) \theta_{\Upsilon}^2 ~.
\end{align}
In terms of the mass eigenstates of the physical particles $\hat{A}_d$ and $\hat{\Upsilon}_d$, the BSEFT Lagrangian reads
\begin{align}\label{eq:bseft-dark-photon-mass-es}
\mathcal{L}_{\rm BSEFT}^{(4)} = & - \frac{1}{4} \hat{F}_d^{\mu\nu}\hat{F}_{d,\mu\nu} - \frac{1}{4} \hat{\Upsilon}_d^{\mu\nu}\hat{\Upsilon}_{d,\mu\nu}\\\nonumber
& + \frac{1}{2}\partial_\mu \eta_d \partial^\mu \eta_d - \frac{m_{\eta_d}^2}{2}\eta_d\eta_d + \frac{\hat{m}_{\Upsilon_d}^2}{2}\hat{\Upsilon}_d^\mu \hat{\Upsilon}_{d,\mu} + \frac{\hat{m}_d^2}{2} \hat{A}_d^\mu \hat{A}_{d,\mu} ~.
\end{align}
Through dark-sector mixing, the dark bound state $\hat{\Upsilon}_d$ inherits all couplings of the dark photon and the $Z$ boson, suppressed by $g_d \epsilon_d$ and $\epsilon t_w g_d\epsilon_d$, respectively. Moreover, through hypercharge mixing, the dark photon $\hat{A}_d$ inherits all interactions of the photon, see~\eqref{eq:field-rotations-fv-model}. In the limit of small hypercharge mixing and dark-sector mixing, the couplings of $\hat{\Upsilon}_d$ to SM fermions are
\begin{align}\label{eq:fv-fermion-interactions}
\mathcal{L} & = - e\epsilon g_d\epsilon_d\frac{m_{\Upsilon_d}^2 - 2m_d^2}{m_{\Upsilon_d}^2 - m_d^2}\hat{\Upsilon}_d^\mu \sum_f Q_f \bar{f}\gamma_\mu f\\\nonumber
& \quad - e\epsilon t_w g_d \epsilon_d\frac{2 m_Z^2 - m_{\Upsilon_d}^2}{m_Z^2 - m_{\Upsilon_d}^2}\hat{\Upsilon}_d^\mu\sum_f\bar{f} \gamma_\mu\left(t_w Q_f P_R - \frac{T_f^3 - s_w^2 Q_f}{s_w c_w}P_L\right)f + \mathcal{O}(\epsilon^2\epsilon_d,\epsilon\epsilon_d^2) ~,
\end{align}
where $T_f^3$ is the weak isospin component of fermion $f$ and $P_R,P_L$ project onto right- and left-chiral fermions.

The mixing-induced interactions from~\eqref{eq:fv-fermion-interactions} generates the decay $\hat{\Upsilon}_d \to f\bar{f}$. Similar to the $\mbox{F}_{\rm S}$ model, we employ the projection techniques from NRQCD~\cite{Petrelli:1997ge} and calculate the bound-state decay from dark fermion annihilation through $\chi\bar{\chi} \to f\bar{f}$ via an off-shell dark photon. In what follows, we neglect small corrections of $\mathcal{O}(\theta_Z^2,\theta_{\Upsilon}^2)$ from vector mixing in the masses and couplings.
 
For small kinetic mixing $\epsilon$ and neglecting $Z-\Upsilon_d$ mixing, we obtain the decay rate
\begin{align}
    \Gamma(\hat{\Upsilon}_d \to f\bar{f}) = \frac{(g_d\,\epsilon e\, Q_f)^2}{12\pi^2}|R_{\Upsilon_d}(0)|^2\left(1 - \frac{4 m_f^2}{m_{\Upsilon_d}^2}\right)^{1/2}\frac{m_{\Upsilon_d}^2 + 2 m_f^2}{(m_{\Upsilon_d}^2 - m_d^2)^2} ~.
\end{align}
By calculating the same decay width in the BSEFT using the first term in~\eqref{eq:fv-fermion-interactions}, we identify the $A_d - \Upsilon_d$ mixing parameter as
\begin{align}\label{eq:eps-d}
\epsilon_d = \frac{R_{\Upsilon_d}(0)}{\sqrt{\pi}m_{\Upsilon_d}^{3/2}}\left(1 - \frac{2 m_d^2}{m_{\Upsilon_d}^2}\right)^{-1}.
\end{align}

Including the $Z-\Upsilon_d$ mixing, the total fermionic decay width is 
\beqa
\Gamma(\hat{\Upsilon}_d \to f\bar{f})= \frac{1}{12\,\pi}\,\left(1 - \frac{4 m_f^2}{m_{\Upsilon_d}^2}\right)^{1/2}\,\frac{\left[g_V^2(m_{\Upsilon_d}^2 + 2 m_f^2) + g_{A}^2(m_{\Upsilon_d}^2 - 4 m_f^2)\right]}{m_{\Upsilon_d}} ~,
\eeqa
with the vector and axial-vector couplings
\beqa
g_V &\equiv& e\epsilon g_d\epsilon_d\frac{m_{\Upsilon_d}^2 - 2m_d^2}{m_{\Upsilon_d}^2 - m_d^2}\,Q_f + e\epsilon t_w g_d\epsilon_d\frac{m_{\Upsilon_d}^2 - 2m_Z^2}{m_{\Upsilon_d}^2 - m_Z^2}\,\left( t_w\,Q_f - \frac{T_f^3}{2 s_w c_w} \right) \\\nonumber
g_A &\equiv&  e\epsilon t_w g_d\epsilon_d\frac{m_{\Upsilon_d}^2 - 2m_Z^2}{m_{\Upsilon_d}^2 - m_Z^2}\,\left( \frac{T_f^3}{2 s_w c_w} \right)\,.
\eeqa

If the dark photon is the mediator of the binding force, the wave function at origin  scales as $|R_{\Upsilon}(0)| \propto \alpha_d^{3/2}$, see~\eqref{eq:wave-origin-fit}. For perturbative couplings $\alpha_d \lesssim 0.5$, the wave function is thus suppressed. Moreover, the bound-state condition~\eqref{eq:bs-condition-s} together with the perturbativity condition $\alpha_d < 0.5$ implies a mild mass hierarchy $m_d <0.15\, m_{\Upsilon_d}$. This means that both $\epsilon_d$ and the mass mixing $\theta_{\Upsilon}$ from~\eqref{eq:fv-mass-mixing} are small. \emph{A posteriori}, neglecting corrections of $\mathcal{O}(\epsilon_d^2)$ and $\mathcal{O}(\theta_{\Upsilon}^2)$ in the field transformations is thus justified.

At mass dimension 5, kinetic mixing can also induce couplings to the neutral pion through the chiral anomaly in QCD,\footnote{If $m_{\eta_d} \approx m_\pi$, then $\eta_d$ can mix with the pion and couplings analogous to~\eqref{eq:pion-couplings} arise.}
\begin{align}\label{eq:pion-couplings}
   \mathcal{L} = & - \frac{\alpha}{4\pi}\frac{\pi^0}{f_\pi}\left(F^{\mu\nu} \widetilde{F}_{\mu\nu} - 2\epsilon F^{\mu\nu} \widetilde{\hat{F}}_{d,\mu\nu} + 2\epsilon g_d\epsilon_d\frac{m_{\Upsilon_d}^2 - 2m_d^2}{m_{\Upsilon_d}^2 - m_d^2} F^{\mu\nu} \widetilde{\hat{\Upsilon}}_{d,\mu\nu} \right) + \mathcal{O}(\epsilon^2)\ ,
\end{align}
where $f_\pi$ is the pion decay constant. The third term induces $\hat{\Upsilon}_d \to \pi^0\gamma$ decays. Since the interaction strength is fixed by~\eqref{eq:eps-d}, we can directly calculate the decay width for this process in the BSEFT. We obtain
\begin{align}
\Gamma(\hat{\Upsilon}_d \to \pi^0\gamma) & = \frac{\alpha_d\, \epsilon^2}{3\pi^3}\frac{\alpha^2}{f_\pi^2}\frac{|R_{\Upsilon_d}(0)|^2}{m_{\Upsilon_d}^3}\left(\frac{m_{\Upsilon_d}^2}{m_{\Upsilon_d}^2 - m_d^2}\right)^2 |\vec{p}|^{3}(m_{\Upsilon_d},m_\pi,0)\ .
\end{align}
with the final-state particle momentum in the $\hat{\Upsilon}_d$ rest frame,
\begin{align}\label{eq:momentum}
    |\vec{p}|(m_a,m_b,m_c) = \frac{1}{2 m_a}\Big[(m_a^2 - (m_b + m_c)^2)(m_a^2 - (m_b - m_c)^2)\Big]^\frac{1}{2}\ .
\end{align}

Up to here, we have only seen interactions of the vector state $\hat{\Upsilon}_d$. Indeed, at mass dimension 4 in the BSEFT, the pseudo-scalar $\eta_d$ does not interact. Due to its quantum numbers, $\eta_d$ does not mix with the dark photon and can only interact through higher-dimensional effective couplings. At mass dimension 5, possible $C_d$- and $P_d$-conserving effective interactions
\begin{align}\label{eq:fv-dim5}
\mathcal{L}_{\rm BSEFT}^{(5)} & = g_d^2\,\frac{C_{\eta_d}}{\Lambda_d}\,\eta_dF_d^{\mu\nu}\widetilde{F}_{d,\mu\nu} + g_d\,\frac{C_{\eta_d\Upsilon_d}}{\Lambda_d}\,\eta_dF_d^{\mu\nu}\widetilde{\Upsilon}_{d,\mu\nu} + \frac{C'_{\eta_d\Upsilon_d}}{\Lambda_d}\eta_d\Upsilon_d^{\mu\nu}\widetilde{\Upsilon}_{d,\mu\nu} ~.
\end{align}
The third term is not relevant for the collider phenomenology we consider, since the decay $\eta_d \to \hat{\Upsilon}_d \hat{\Upsilon}_d$ is kinematically forbidden.
The first interaction term induces the decay $\eta_d \to \gamma_d\gamma_d$. To determine the BSEFT coefficient $C_{\eta_d}$, we calculate the amplitude for $\eta_d \to \gamma_d \gamma_d$ using projection techniques and compare the result with the same amplitude calculated based on~\eqref{eq:fv-dim5}. We obtain the relation
 \begin{align}\label{eq:C-eta}
      \frac{C_{\eta_d}(p_1,p_2)}{\Lambda_d} = \frac{2R_{\eta_d}(0)}{\sqrt{\pi\,m_{\eta_d}}} \frac{1}{(p_2 - p_1)^2 - m_{\eta_d}^2} ~,
 \end{align}
where $p_1$ and $p_2$ are the four-momenta of the outgoing dark photons, which must fulfill the condition $|\vec{p}_1 - \vec{p}_2| \ll m_{h_d}$, as the dark scalars in~\eqref{eq:lambdah}. Notice that \eqref{eq:fv-dim5} together with \eqref{eq:C-eta} applies for generic interactions of an on-shell $\eta_d$ with on-shell or off-shell dark photons.

In the decay rate for $\eta_d\to \gamma_d\gamma_d$, the momenta $p_1$ and $p_2$ are fixed by the two-body kinematics in the rest frame of the bound state. At the leading order in $\alpha_d$, we calculate the decay rate in the BSEFT as
\begin{align}\label{eq:etad-to-gammad-gammad}
 \Gamma(\eta_d\to \gamma_d\gamma_d) = 4\frac{\alpha_d^2}{m_{\eta_d}^2} |R_{\eta_d}(0)|^2 \left(1 - \frac{4 m_d^2}{m_{\eta_d}^2}\right)^{\frac{3}{2}} \left(1 - \frac{2 m_d^2}{m_{\eta_d}^2}\right)^{-2}.
\end{align}
This decay rate corresponds to that for para-positronium~\cite{Berko:1980gg}  using the Coulomb wave function, $|R_{\eta_d}(0)|^2 = (\alpha_d m_\chi)^3/2$, and taking $\alpha_d \to \alpha$, $m_{\eta_d} \to 2m_e$ and $m_d \to 0$.

The second interaction term in~\eqref{eq:fv-dim5} induces the decay $\Upsilon_d \to \eta_d\gamma_d$. This decay is induced by a magnetic dipole transition and can be calculated in NRQCD~\cite{Brambilla:2005zw}. We find the decay rate
\begin{align}\label{eq:fv-ups-to-eta-gammad}
\Gamma(\Upsilon_d \to \eta_d\gamma_d) = \frac{4\alpha_d}{3}\,\frac{1}{m_\chi^2}|\vec{p}|^{3}(m_{\Upsilon_d},m_{\eta_d},m_d) \approx \frac{2\alpha_d}{3}\frac{(m_{\Upsilon_d}^2 - m_{\eta_d}^2)^3}{m_{\Upsilon_d}^5}\ .
\end{align}
For small mass splitting $\Delta = (m_{\Upsilon_d}^2 - m_{\eta_d}^2)/m_{\Upsilon_d}^2 \ll 1$ and $m_d \ll m_{\Upsilon_d} - m_{\eta_d}$, the decay rate is suppressed as $\Delta^3$. By matching the amplitude onto the BSEFT result, we identify the effective coupling in~\eqref{eq:fv-dim5} as
\begin{align}
    \frac{C_{\eta_d\Upsilon_d}}{\Lambda_d} = \frac{1}{2m_\chi} ~.
\end{align}
Notice that~\eqref{eq:fv-ups-to-eta-gammad} applies for weakly coupled dark sectors. If the binding force is strongly coupling, the non-perturbative contributions to the $\Upsilon_d \to \eta_d$ transition can be described by a form factor.

Due to the phase-space suppression of $\Upsilon_d \to \eta_d\gamma_d$ decays, the three-body decay $\Upsilon_d \to 3\gamma_d$ dominates the total decay width in most of the parameter space. For massless dark photons, the decay rate for $\Upsilon_d \to 3\gamma_d$ is analogous to orthopositronium decays in QED. At leading order in $\alpha_d$, we deduce~\cite{Ore:1949te,Adkins:1996odo}
\begin{align}
    \Gamma(\Upsilon_d \to 3 \gamma_d) = \frac{16(\pi^2 - 9)}{9\pi m_{\Upsilon_d}^2}\alpha_d^3\, |R_{\Upsilon_d}(0)|^2\ .
\end{align}
The result for massive dark photons can be found in~\cite{An:2015pva}.

Using the matching conditions derived above, we can specify the general BSEFT interactions~\eqref{eq:bseft-dark-photon} and~\eqref{eq:fv-dim5} for the case that the binding force is the mediator to the Standard Model. The relevant BSEFT terms read
\begin{align}
    \mathcal{L}_{\rm BSEFT} = & \ -\frac{\epsilon g_d}{\sqrt{\pi}}\,\frac{R_{\Upsilon}(0)\sqrt{m_{\Upsilon_d}}}{m_{\Upsilon_d}^2 - m_d^2}\bigg(\hat{\Upsilon}_d^\mu \sum_f e\,Q_f \bar{f}\gamma_\mu f + \frac{\alpha}{2\pi f_\pi}\,\pi^0 F^{\mu\nu} \widetilde{\hat{\Upsilon}}_{d,\mu\nu}\bigg)\\\nonumber
    & + \frac{g_d^2}{\sqrt{4\pi m_{\eta_d}}}\,\frac{R_{\eta}(0)}{(p_2 - p_1)^2-m_{\eta_d}^2}\,\eta_d\, F_d^{\mu\nu}\widetilde{F}_{d,\mu\nu} + \frac{g_d}{m_{\Upsilon_d}}\,\eta_d F_{d}^{\mu\nu} \widetilde{\hat{\Upsilon}}_{d,\mu\nu}\ ,
\end{align}
where $p_1$ and $p_2$ are the four-momenta of the outgoing dark photons and we have used $2 m_\chi \approx m_{\Upsilon_d}$ in the last term.

We stress once more that the BSEFT Lagrangian can be used to calculate arbitrary processes with on-shell bound states in relativistic field theory. The BSEFT expansion in the relative velocity $v$ of the bound-state constituents $\chi$ ensures that their relative momenta are small compared to the bound-state mass $M$ and the bound-state conditions are satisfied. Moreover, using the BSEFT for scattering processes automatically factorizes low-energy from high-energy contributions to the transition amplitude. This factorization is apparent in the momentum-dependent BSEFT coefficients $\lambda_h(p_1,p_2)$ from~\eqref{eq:lambdah} and $C_{\eta_d}(p_1,p_2)$ from~\eqref{eq:C-eta}, which fulfill the condition $|\vec{p}_1 - \vec{p}_2| \ll M$, but allow for $|\vec{p}_1 + \vec{p}_2| \gg M$.

\section{Collider phenomenology of darkonia}
\label{SEC:collider}
The BSEFT is a convenient calculation framework for darkonium phenomenology. In this section, we apply the BSEFT to make predictions for darkonium production and decay at colliders. We focus on new signatures at the LHC and at Belle II. For sub-GeV dark sectors, experiments with a long baseline and/or a high-luminosity particle source are interesting alternatives. In particular, the far-distance experiment FASER~\cite{FASER:2018eoc} at CERN or the fixed-target experiments NA62~\cite{NA62:2017rwk}, NA64~\cite{NA64:2016oww,NA64:2020qwq} and the future SHiP experiment~\cite{SHiP:2021nfo} can probe darkonia with tiny interactions with the Standard Model.

At colliders, darkonia can be efficiently produced through production processes of SM particles they mix with. For instance, scalar darkonium can mix with the Higgs boson and be produced from Higgs production or decay. In general, a sizeable darkonium production rate through mixing requires a sizeable dark coupling strength $\alpha_d$. If the mediator force is identical to the binding force and $\alpha_d$ is large, the bound-state condition $m_d \lesssim \alpha_d m_\chi/2$ implies that the mediator should be lighter than the darkonium state. In general, the mediator to the Standard Model can be independent of the binding force and no such mass hierarchy is imposed.

Dark bound states can decay either into SM particles through mixing with the mediators, or into pairs of mediators or lighter bound states through dark-sector interactions. In the minimal scenario where the binding force is identical to the mediator force, the branching ratio for these decay channels is well-defined. In general, decays into dark-sector particles depend on the binding force and additional model assumptions are needed to determine the partial decay rates. In our analysis, we make concrete numerical predictions for signatures in the minimal scenario and discuss the general scenario at a qualitative level.

The darkonium models we consider in this work predict a large variety of collider signatures. In general, a darkonium state is produced and decays, possibly through a cascade, to SM particles. Alternatively, a stable darkonium state can leave the detector unseen. Depending on the masses and couplings of the mediator particles, the final-state decay products can appear prompt, displaced or invisible to the detector. Since darkonium production favors sizeable dark couplings $\alpha_d$, the darkonium decay to mediators is typically prompt, unless another suppression mechanism is at work. Light mediators with small couplings to SM particles typically decay at a displacement from the production point, which can be reconstructed as a displaced vertex of two charged tracks. Signatures of invisible particles can arise from stable darkonia or force carriers, or from mediators with a long decay length compared to the detector scales.

In what follows, we describe the collider phenomenology for each model, focusing on bound-state masses from a few GeV up to several hundred GeV. Weak-scale darkonia have been discussed before, mostly in the context of supersymmetry, as in Refs.~\cite{Nappi:1981ft,Barger:1988sp,Martin:2008sv}.

\subsection{$\mbox{F}_{\rm S}$ model}
\label{sec:fs-pheno}
The phenomenology of the $\mbox{F}_{\rm S}$ model is determined by the masses $m_{h_d}$, $m_{\eta}$, $m_{\Upsilon}$, $m_S$, several dark-sector couplings and the mixing angles $\theta_h$ and $\theta_{h_d}$. We assume the mass hierarchy $m_{h_d} > m_{\Upsilon_d} = m_{\eta_d} > m_S$. Choosing $h_d$ as the heaviest bound state is motivated by phenomenology, since $\Upsilon_d$ and $\eta_d$ are stable and thus invisible. We neglect the hyperfine splitting between $\Upsilon_d$ and $\eta_d$, which leaves the phenomenology untouched, since these states cannot decay into each other.

\paragraph{Decay}
The decay channels and branching ratios of the scalar darkonium state $h_d$ are determined by its interactions in~\eqref{eq:fs-model-interactions}. For small mass mixing, the dominant decay modes are
\begin{align}\label{eq:fs-model-hd-decays}
    h_d  \stackrel{\lambda_h}{\longrightarrow} SS \ ,\qquad
    h_d  \stackrel{\xi_{\eta_d}}{\longrightarrow} \eta_d \eta_d  \ , \qquad
    h_d  \stackrel{C_{h\Upsilon}}{\longrightarrow} \Upsilon_d \Upsilon_d \ .
\end{align}
The corresponding decay rates are given in~\eqref{eq:hd-to-s-s}, \eqref{eq:hd-to-etad-etad} and \eqref{eq:hd-to-upsd-upsd}. The relative size of these decay rates is difficult to estimate, due to the unknown $h_d \eta_d\eta_d$ and $h_d \Upsilon_d\Upsilon_d$ couplings, see Sec.~\ref{sec:fs-model}. We therefore treat the relative decay rates as free parameters compared to the calculable partial rate for $h_d \to SS$.

Due to the assumed $P_d$ and $C_d$ symmetries in the dark sector, the processes in~\eqref{eq:fs-model-hd-decays} cover all possible options for $h_d$ to decay into dark-sector particles. Further decay channels are suppressed by mass mixing $\theta_h$ and $\theta_{h_d}$. In particular, the $h_d$ decay rate to SM particles is suppressed as $\theta_h^2\theta_{h_d}^2$.

The scalar mediator $S$ inherits the decay modes of the Higgs boson, suppressed by scalar mixing. Possible decay modes are $S \to \ell^+ \ell^-,\, jj,\, \gamma\gamma$, with the branching ratios determined by the relative Higgs couplings to these states~\cite{Winkler:2018qyg}. All decay rates of $S$ are suppressed by $\theta_h^2$. For small Higgs mixing $\theta_h$ and $m_S \lesssim$ GeV, the decays are likely to occur with a displacement from the production point.

\paragraph{Production through Higgs decays}
Since the interactions of all darkonia with the Standard Model are suppressed by mass mixing, dark bound states are mostly produced through the scalar mediator $S$ via the Higgs portal couplings $\mu_S$ and $\lambda_S$. However, since $S$ is typically the lightest state of the dark sector, the darkonia are not produced from resonant $S$ decays, but through mass mixing or off-shell mediators.

At the LHC, dark scalars can be produced through Higgs decays
\begin{align}\label{eq:fs-model-higgs-decays}
    pp & \to h \to SS  & \sim & \ \mu_S\theta_h + \lambda_S v\\\nonumber
    pp & \to h \to Sh_d  & \sim & \ \lambda_h \theta_h - \lambda_S v\, \theta_{h_d}\\\nonumber
    pp & \to h \to h_dh_d  & \sim & \ \lambda_h' \theta_h\ .
\end{align}
The scaling with the model parameters follows from the couplings in~\eqref{eq:fs-model-interactions}. 
The partial decay widths of the Higgs boson into dark scalars can be calculated from the results in Sec.~\ref{sec:fs-model}. We obtain
\begin{align}\label{eq:higgs-partial-decays}
\Gamma(h \to SS) & = \frac{|\mu_S \theta_h + \lambda_S v|^2}{32\pi m_h}\bigg(1 - \frac{4 m_S^2}{m_h^2}\bigg)^{\frac{1}{2}}\\\nonumber
\Gamma(h \to Sh_d) & = \frac{|\lambda_h(p_h,p_S) \theta_h - \lambda_S v\, \theta_{h_d}|^2}{2\pi m_h^2} |\vec{p}|(m_h,m_S,m_{h_d}) \propto |R_{h_d}'(0)|^2\\\nonumber
\Gamma(h \to h_dh_d) & = \frac{|\lambda_h' \theta_h|^2}{32\pi m_h}\bigg(1 - \frac{4 m_{h_d}^2}{m_h^2}\bigg)^{\frac{1}{2}} \propto |R_{h_d}'(0)|^4\ ,
\end{align}
with the final-state momentum $|\vec{p}|(m_h,m_S,m_{h_d})$ defined in~\eqref{eq:momentum} and the BSEFT coupling for the process $h \to Sh_d$ (see~\eqref{eq:lambdah})
\begin{align}
    \lambda_h(p_h,p_S) = 64\sqrt{\pi} \alpha_d\frac{R_{h_d}'(0)}{m_{h_d}^{3/2}} \frac{m_h^2\,|\vec{p}|^2(m_h,m_S,m_{h_d})}{(m_h^2 - m_{h_d}^2 + m_S^2)^2}\ .
\end{align}
The relative magnitude of the three decay rates is determined by the model parameters $\lambda_S$, $\theta_h \propto \mu_S$, $\{\theta_{h_d},\lambda_h\} \propto R_{h_d}'(0)$ and $\lambda_h' \propto |R_{h_d}'(0)|^2$.

The sum of the three partial decay widths, $\Gamma_{\rm NP} = \Gamma(h \to SS) + \Gamma(h \to Sh_d) + \Gamma(h \to h_dh_d)$, can be constrained from combined measurements of Higgs production and decay rates to SM final states. Such an approach makes no assumption about the contributions to additional decay modes, as long as the final states are not too similar to the SM decay modes. In the $\mbox{F}_{\rm S}$ model, this is indeed the case, since the masses, decay lengths and topologies of the involved particles generally differ from the Standard Model prediction.

In Ref.~\cite{Biekotter:2022ckj}, the authors have constrained the branching ratio of such new Higgs decays, $\text{BR}_{\rm inv}$, from a large set of Higgs measurements in LHC data. Under the assumption that the Higgs couplings to SM particles are scaled by a universal mixing parameter $\cos\theta$, they derive the upper bound for small $\theta$~\cite{Biekotter:2022ckj}
\begin{align}\label{eq:BRinv}
   \text{BR}_{\rm inv} < 0.078\left(1 - \left(\frac{\theta}{0.285}\right)^2\right)\qquad \text{at } 95\%\,\text{C.L.}
\end{align}
In the $\mbox{F}_{\rm S}$ model, $\text{BR}_{\rm inv} = \Gamma_{\rm NP}(\theta_h)/(\Gamma_{\rm SM} + \Gamma_{\rm NP}(\theta_h))$ with $\Gamma_{\rm SM} = 4.1\,$MeV, where $\theta_h$ corresponds with the mixing angle $\theta$.

Let us have a closer look at the three decay modes~\eqref{eq:higgs-partial-decays}. For bound-state production via $h \to Sh_d$ to dominate over $h \to SS$, the quartic Higgs portal coupling $\lambda_S$ should be small. Moreover, the BSEFT coupling $\lambda_h$ should be larger than the Higgs mixing parameter $\mu_S$, which implies $\lambda_h \propto R_{h_d}'(0) \propto (\alpha_d m_\chi)^{5/2} > \mu_S$. In what follows, we set $\lambda_S = 0$ and focus on the parameter region with $\lambda_h > \mu_S$. In this case, all three decay modes are proportional to $\theta_h^2$.

In the minimal scenario, searches for $B \to K S$ decays at flavor physics experiments strongly constrain the Higgs mixing angle $\theta_h < 10^{-3} - 10^{-4}$ for $m_S \lesssim 4.5\,$GeV~\cite{Batell:2022dpx}. This constraint suppresses the Higgs decay rates~\eqref{eq:higgs-partial-decays} well below the SM decay rate. For $m_S \gtrsim 4.5\,$GeV, $\theta_h$ is only subject to the bound from Higgs physics~\eqref{eq:BRinv}. However, the bound-state condition~\eqref{eq:bs-condition-p} sets a lower bound on the bound-state mass, $m_{h_d} > 4\,m_S/(0.22\,\alpha_d)$. Bound-state production from Higgs decays is viable only if $m_{h_d}$ lies not far below the Higgs mass and if the coupling $\alpha_d$ is strong.

\begin{figure}[t!]
\centering
    \includegraphics[width=0.48\textwidth]{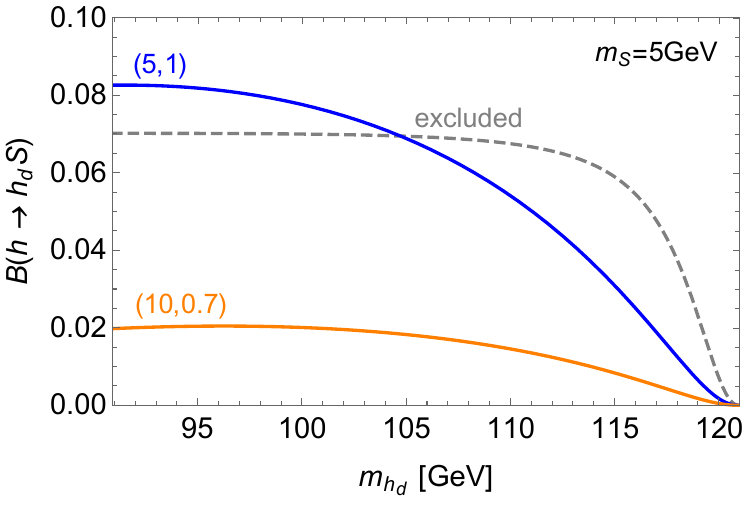} \hfill
    \includegraphics[width=0.48\textwidth]{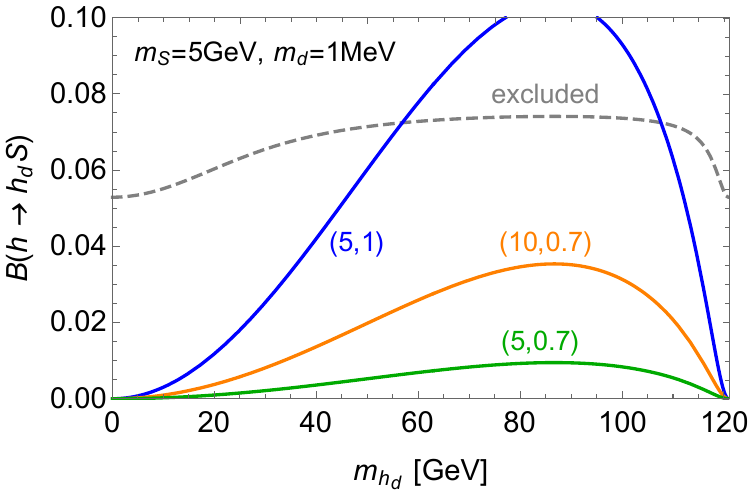} 
    \caption{Branching ratio for $h \to h_d S$ as a function of the bound-state mass $m_{h_d}$ and for pairs of $(\mu_S [\text{GeV}],\alpha_d)$. The Higgs-portal coupling is fixed to $\lambda_S = 0$. The area above the dashed gray curve is excluded by Higgs coupling measurements~\eqref{eq:BRinv} for $\alpha_d \le 1$. Left: Minimal scenario. The model parameters $m_{h_d}$, $m_S$ and $\alpha_d$ fulfill the bound-state condition~\eqref{eq:bs-condition-p}. Right: The dark scalar is not associated with the binding force, but has the same coupling to dark fermions.
    \label{fig:FS-decay}}
\end{figure}

In Fig.~\ref{fig:FS-decay}, left, we show the branching ratio for $h \to h_d S$ in the minimal scenario as a function of the bound-state mass $m_{h_d}$ for two benchmarks $(\mu_S,\alpha_d)$. The branching ratio scales with these parameters as roughly $\mathcal{B}(h \to h_d S) \propto \alpha_d^{5/2} \mu_S^4$. The dashed curve shows the maximum branching ratio that is in agreement with the bound from Higgs coupling measurements~\eqref{eq:BRinv} for fixed $\alpha_d = 1$ and $m_S = 5\,$GeV.

In general, the mediator $S$ does not have to be associated with the binding force, so that $m_d \neq m_S$. In this case, the mass and coupling of the binding force, $m_d$ and $\alpha_d$, only enter through the wave-function derivative $R'(0)$, see~\eqref{eq:fs-derivative}. For $m_d \ll \alpha_d m_{h_d}$, the bound-state condition is easily satisfied and $m_{h_d}$ is essentially a free parameter. The bound-state mixing with the mediator still needs to be strong to ensure that $h \to Sh_d$ dominates over $h \to SS$. For light mediators with $m_S \ll m_h$, the Higgs decay phenomenology is then largely independent of $m_S$.

In Fig.~\ref{fig:FS-decay}, right, we display the branching ratio for $h \to h_d S$ in the general scenario for three benchmarks $(\mu_S,\alpha_d)$. To determine the wave-function derivative $R'(0)$, we have assumed that the binding force carrier has mass $m_d = 1\,$MeV and the same coupling strength as the mediator $S$. Increasing the coupling $\alpha_d$ (the mass $m_d$) means increasing (decreasing) $R'(0)$, see Fig.~\ref{fig:bindingenergy-p-wave}, and thereby increasing (decreasing) the branching ratio. Taking this variation of the wave function into account, the branching ratios in Fig.~\ref{fig:FS-decay}, right, can be used to estimate the expected production rates of generic dark-fermion bound states which interact with the Standard Model through a Higgs-portal scalar.

\begin{figure}[t!]
\centering
    \includegraphics[width=0.6\textwidth]{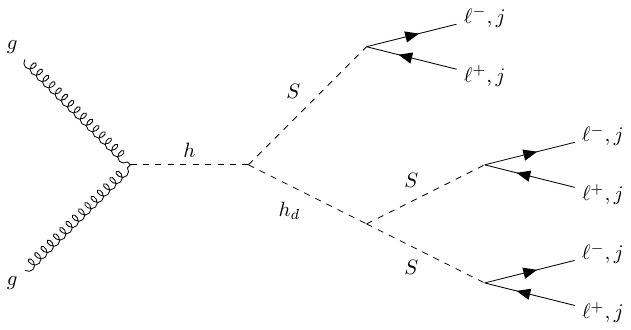} 
    \caption{Feynman diagram for $h_d + S$ production from the Higgs boson decay with a large multiplicity of SM particles in the final state. For light scalars, the decay $S\rightarrow \gamma\gamma$ can also be relevant.
    \label{fig:FS-Feynman1}}
\end{figure}

\paragraph{Signatures from Higgs decays} At the LHC, the $\mbox{F}_{\rm S}$ model predicts the following dominant darkonium signatures:
\begin{align}\label{eq:FS-higgs-sig}
    pp & \to h \rightarrow S h_d \to [\text{SM}]_{S} [SS]_{h_d} \to [\text{SM}]_{S}\big[[\text{SM}]_{S} [\text{SM}]_{S}\big]_{h_d}\ ,
\end{align}
where $\text{SM} = \{\ell^+\ell^-,jj,\gamma\gamma\}$. The four-momenta of the particles inside the brackets $[\dots]_x$ reconstruct the mass of the mother particle $x$. The Feynman diagram is shown in Fig.~\ref{fig:FS-Feynman1}.

For the minimal scenario, where the mediator to the Standard Model is identical to the binding force, we define the benchmark
\begin{align}
m_S = 5\,\text{GeV}\ ,\ m_{h_d} = 105\,\text{GeV}\ ,\ \mu_S = 5\,\text{GeV}\ ,\ \alpha_d = 1\ ,\ \lambda_S = 0\ .
\end{align}
For these parameters, the Higgs branching ratio into bound states is $\mathcal{B}(h \to h_d S) = 0.069.$ To predict the bound-state production rate at the 14-TeV LHC, we use the Higgs production cross section for vector boson fusion (VBF), $4277\,$fb, corresponding to a di-jet invariant mass of $m_{jj} = 130\,$GeV~\cite{LHCHiggsCrossSectionWorkingGroup:2016ypw}. The cross section for bound-state production via $pp\to hjj \to h_dSjj$ is then $326\,$fb.

In the general scenario, the $h_d$ production cross section is similar in magnitude compared to the minimal scenario. The main difference is that bound states $h_d$ much lighter than the Higgs boson can be produced, as long as the bound-state condition $m_{h_d} > 4 m_d/(0.22\alpha_d)$ is fulfilled.

In both scenarios, the signature~\eqref{eq:FS-higgs-sig} consists of three pairs of SM particles, where two of these pairs reconstruct the $h_d$ mass. For light mediators $S$ below the GeV scale, the decay can occur with a displacement from the production point.

In the general scenario, the bound state $h_d$ has an extra decay mode into binding force carriers, which competes with the $h_d \to SS$ decay. If the force carrier does not interact with the Standard Model, the bound-state decay appears invisible in the detector. This results in the signature
\begin{align}
pp & \to h \rightarrow S h_d \to [\text{SM}]_{S} +  \slashed{E}_T\ ,
\end{align}
where $\slashed{E}_T$ stands for missing transverse energy.

\paragraph{Production from gluon-gluon fusion}

Alternatively to Higgs decays, the scalar bound state $h_d$ can be produced directly from proton-proton collisions through scalar mixing. Direct production is mostly relevant for darkonia heavier than the Higgs boson. For lighter darkonia, the expected event rate from Higgs decays is higher than from direct production.

The total cross section for $h_d$ production from gluon-gluon fusion is given by
\begin{align}\label{eq:hd-gg}
    \sigma(pp\to h_d) = \sigma(pp \to h)_{m_{h_d}} \sin^2\theta_h\sin^2\theta_{h_d}\ ,
\end{align}
where $\sigma(pp \to h)_{m_{h_d}}$ denotes the Higgs production cross section from gluon-gluon fusion for a certain mass parameter $m_{h_d}$. Darkonium production from VBF or in association with a $Z$ boson are viable alternatives.

For darkonia heavier than the Higgs boson, the branching ratio for new Higgs decays, $\text{BR}_{\rm inv}$, is saturated by $h \to SS$. The bound from Higgs coupling measurements~\eqref{eq:BRinv} translates into an upper bound on the Higgs mixing parameter, $\mu_S \lesssim 10\,$GeV for mediator masses $m_S \ll m_h$, which corresponds to a $S - h$ mixing angle $\theta_h \lesssim 0.15$. For $m_{h_d} > m_h$, this bound is independent of $m_{h_d}$ and $\alpha_d$, since $\text{BR}_{\rm inv}$ is saturated by $h\to SS$. In addition, $S-h_d$ mixing is suppressed by the wave-function derivative, yielding $\theta_{h_d} \lesssim 0.08$ for $\alpha_d \le 1$ and $m_S \ll m_{h_d}$. These constraints suppress the $h_d$ production cross section significantly.

\begin{figure}[t!]
\centering
    \includegraphics[width=0.485\textwidth]{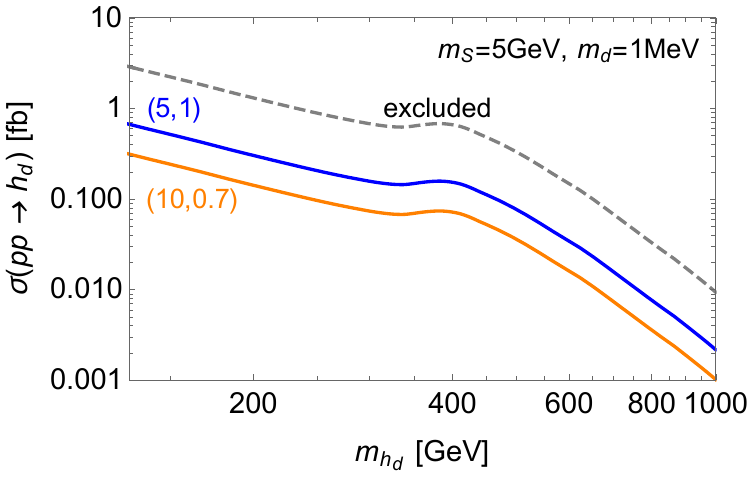} \hfill
    \includegraphics[width=0.48\textwidth]{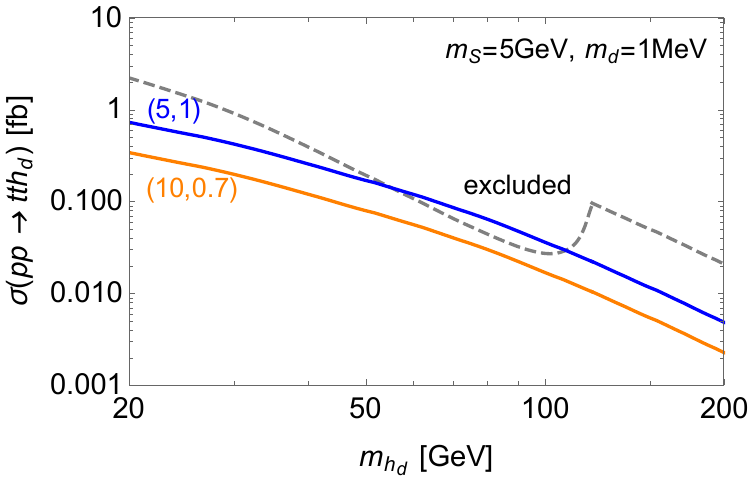} 
    \caption{Darkonium production cross sections at the 14-TeV LHC as a function of the bound-state mass $m_{h_d}$ and for pairs of $(\mu_S [\text{GeV}],\alpha_d)$ in the general scenario. Left: $h_d$ production from gluon-gluon fusion. Right: $h_d$ production in association with a top-antitop pair. The area above the dashed gray curve is excluded by Higgs measurements~\eqref{eq:BRinv} for $\alpha_d \le 1$.
    \label{fig:FS-hd-gg}}
\end{figure}

In Fig.~\ref{fig:FS-hd-gg}, left, we show the $h_d$ production cross section~\eqref{eq:hd-gg} at the 14-TeV LHC as a function of the darkonium mass. The prediction is based on the leading-order Higgs production cross section for a variable scalar mass from Ref.~\cite{Djouadi:2005gi}. For darkonia not much heavier than the Higgs boson, the cross section ranges around a few femtobarns.

\paragraph{Signatures from gluon-gluon fusion} 
 The signatures expected from direct $h_d$ production are
\begin{align}
pp \to h_d \to [SS]_{h_d} \to \big[[\text{SM}]_{S} [\text{SM}]_{S}\big]_{h_d}\ ,
\end{align}
consisting of a pair of (displaced) SM particles, which together reconstruct the $h_d$ mass. Notice that the branching ratio for $h_d \to SS$ is close to 100\%, since $h_d$ decay rates to SM particles are suppressed by mixing $(\theta_h\theta_{h_d})^2$.

For the general scenario, we consider a benchmark where the darkonium production rate is close to maximal\footnote{The production cross section is largely independent of $\lambda_S$, which only affects the upper bound on the Higgs mixing angle $\theta_h$.} 
\begin{align}
m_S = 5\,\text{GeV}\ ,\ m_{h_d} = 200\,\text{GeV}\ ,\ m_d = 1\,\text{MeV}\ ,\ \mu_S = 10\,\text{GeV}\ ,\ \alpha_d = 1\ .
\end{align}
For this parameter choice, the cross section at the 14-TeV LHC is $\sigma(pp\to h_d) = 1.2\,$fb. In the minimal scenario, the cross section mildly decreases when increasing the binding-force-carrier mass $m_d$.

Similar to Higgs decays, the signature features two (displaced) vertices of SM particles, which together reconstruct the darkonium mass. A recent search by ATLAS with $139/$fb of 13-TeV LHC data has constrained the cross section for a similar process with scalars $S$ decaying into two vector bosons $Z_d$ in $pp\to S \to [Z_d Z_d]_S \to [[\text{SM}]_{Z_d} [\text{SM}]_{Z_d}]_S$~\cite{ATLAS:2024zoq}. For $m_S = 200\,$GeV and $m_{Z_d} = 10\,$GeV, this search constrains the production rate for a four-lepton final state to $\sigma\times\mathcal{B} \lesssim 0.1\,$fb at the 95\% C.L., with a somewhat lower sensitivity for larger $m_S$ masses. This search demonstrates that the darkonium signatures with large decay branching ratios are statistically within reach at the LHC. However, for scalars heavier than a few GeV, the branching ratio into electrons and muons is at the sub-percent level~\cite{Bezrukov:2013fca} and decays into tau leptons and jets dominate. Searches for final states with pairs of taus or pairs of jets in Run-3 data can be sensitive to scalar darkonium in the $F_S$ model.

In the general scenario, the darkonium tends to decay into the binding-force carriers. This can result in signatures with missing energy or (displaced) SM particles, depending on whether or not the force carrier is invisible for detection purposes. For invisible final states, initial-state jet radiation or $h_d$ production through vector boson fusion can be considered for triggering purposes. For visible final states, the decay topology is the same as in the ATLAS analysis~\cite{ATLAS:2024zoq}, but with displaced, rather than prompt, final-state leptons or jets. We encourage a search for two displaced vertices produced through $S \to [Z_d Z_d]_S \to [[\text{SM}]_{Z_d} [\text{SM}]_{Z_d}]_S$, possibly extending to smaller $Z_d$ masses, where displaced decays are most likely to occur.

\paragraph{Production in association with top quarks} Darkonia can also be produced in association with a top-antitop pair, which can be used to trigger on the event. The cross sections for $S$ and $h_d$ production are given by
\begin{align}\label{eq:top-associate}
\sigma(pp \to t\bar{t} S) & = \sigma(pp \to t\bar{t}\phi)_{m_S} \sin^2\theta_h \\\nonumber
\sigma(pp \to t\bar{t} h_d) & = \sigma(pp \to t\bar{t}\phi)_{m_{h_d}} \sin^2\theta_h\sin^2\theta_{h_d}\ .
\end{align}
The cross section for $pp\to t\bar{t}\phi$ with a generic scalar $\phi$ with Yukawa-like couplings to SM fermions has been predicted for the LHC in Refs.~\cite{Haisch:2016gry,Plehn:2017bys}. For light scalars produced at low momenta, the cross section features a soft-collinear enhancement. For heavier scalars, the cross section steeply decreases.

As in gluon-gluon fusion, both cross sections in~\eqref{eq:top-associate} are suppressed by Higgs mixing $\theta_h^2$, which is subject to the bound from Higgs coupling measurements~\eqref{eq:BRinv}. In addition, the bound-state production cross section features an extra suppression by the $S-h_d$ mixing $\theta_{h_d}^2$. For viable parameter combinations, the production rate for $pp\to t\bar{t}h_d$ typically ranges below a few femtobarns. An exception is strong $S-h_d$ mixing for $m_S \approx m_{h_d}$, where the suppression of the cross section by $\theta_{h_d}^2$ is lifted.

In the minimal scenario, the bound-state condition requires $m_S \ll m_{h_d}$ and $t\bar{t}S$ production dominates over $t\bar{t} h_d$, since radiation of the heavier bound state is kinematically suppressed. In Fig.~\ref{fig:FS-hd-gg}, right, we show the cross section for $t\bar{t}h_d$ production in the general scenario, where the darkonium mass is independent of the mediator mass. Notice that the bound on the cross section from Higgs decays (dashed gray curve) is different for $m_{h_d} + m_S < m_h$ and $m_{h_d} + m_S > m_h$, since the decay $h \to h_d S$ is kinematically forbidden for darkonium masses above the Higgs mass.

The signatures from top-associated darkonium production are similar to what has been described above for gluon-gluon fusion.

\paragraph{Production from $B$ meson decays} 
Dark scalars lighter than about 4.5\,GeV can be produced in loop-induced $B$ meson decays by coupling through the Higgs portal to the top quark and $W$ boson in the loop. 
The branching ratios for $X = \{S,h_d\}$ production are given by~\cite{Filimonova:2019tuy}
\begin{align}
    \mathcal{B}(B \to K X) = \frac{\sqrt{2}G_F}{32\pi\Gamma_B}\frac{|C_{bs}^X|^2}{m_B^2}\frac{(m_b + m_s)^2}{(m_b - m_s)^2}f_0^2(m_X^2)(m_B^2 - m_K^2)^2 |\vec{p}|(m_B,m_K,m_X)\ ,
\end{align}
with the Wilson coefficients for the loop-induced $\bar{s}bX$ coupling
\begin{align}
    C_{bs}^S & = \frac{3\sqrt{2}G_F m_t^2}{16\pi^2}V_{tb}V_{ts}^\ast
    \sin\theta_h\\\nonumber
    C_{bs}^{h_d} & = \frac{3\sqrt{2}G_F m_t^2}{16\pi^2}V_{tb}V_{ts}^\ast\sin\theta_h\sin\theta_{h_d}\ .
\end{align}
For equal masses $m_S \approx m_{h_d}$, darkonium production is relatively suppressed by the scalar mixing $\theta_{h_d}$.

In the minimal scenario, $m_S \ll m_{h_d}$ and $B \to K S$ dominates over $B \to K h_d$. For mediator masses above the di-muon threshold, the $S-h$ mixing is strongly constrained by searches for $B \to K S$ with $S \to \mu^+\mu^-$~\cite{Filimonova:2019tuy}. Assuming $\mathcal{B}(B \to K S)\ = 100\%$, a search for displaced di-muons at LHCb~\cite{LHCb:2016awg} yields the bound $\theta_h < 10^{-3}-10^{-4}$ for $m_S < m_B - m_K$. For smaller mediator masses, bounds from searches for rare kaon decays at fixed-target experiments are similar in strength, see~\cite{Batell:2022dpx} for an overview. Taking these constraints into account, the branching ratio for darkonium production is around
\begin{align}
\mathcal{B}(B \to K h_d) \approx 0.5\cdot 10^{-8}\left(\frac{\theta_h}{10^{-4}}\right)^2\sin^2\theta_{h_d}\ .
\end{align}
In the general scenario, the mediator $S$ can be too heavy to be produced in $B$ decays. In this case, the Higgs mixing $\theta_h$ is only constrained from Higgs coupling measurements~\eqref{eq:BRinv}, which are much weaker compared to the bounds from meson decays. The branching ratio for darkonium production for $m_{h_d} < \{m_B - m_K,m_S\}$ can thus be sizeable, reaching up to $\mathcal{B}(B \to K h_d) \approx 10^{-3}$.

\paragraph{Signatures from $B$ meson decays} At Belle II and LHCb, darkonium production in $B$ meson decays leads to the dominant signatures
\begin{align}\label{eq:FS-b-decays}
    m_S < m_{h_d}/2:\quad B & \to K h_d \to K [S S]_{h_d} \to K \big[(\text{SM})_{S}(\text{SM})_{S}\big]_{h_d}\\\nonumber
    m_S > m_{h_d}/2: \quad B & \to K h_d \to K (\text{SM})_{h_d}\ .
\end{align}
The first signature consists of a kaon plus two (displaced) pairs of SM particles, which reconstruct the $h_d$ mass. In the minimal scenario, the bound-state condition~\eqref{eq:bs-condition-p} implies $m_S \lesssim 250\,$MeV for $\alpha_d < 1$. In this case, the signature consists of two lepton pairs, which can be electrons or, if kinematically allowed, muons. For the benchmark scenario
\begin{align}
m_S = 100\,\text{MeV}\ ,\ m_{h_d} = 2\,\text{GeV}\ ,\ \mu_S = 15\,\text{MeV}\ ,\ \alpha_d = 1\ ,
\end{align}
the bounds on $\theta_h$ are fulfilled. The signature consists of two displaced electron pairs from mediator decays, produced with a branching ratio of
\begin{align}
\mathcal{B}(B \to K h_d \to K [S S]_{h_d}) \approx 1.3\cdot 10^{-10}\ .
\end{align}
Despite the small expected event rate, Belle II or LHCb could be sensitive to this new decay topology with a dedicated, essentially background-free analysis.\footnote{For a similar signature, Belle II's sensitivity has been predicted in Ref.~\cite{Cheung:2024oxh}.}

The second signature is only relevant in the general scenario, where the mediator can be heavy and the darkonium decays dominantly into SM particles. The final state then consists of a kaon and one pair of (displaced) SM particles, which reconstruct the $h_d$ mass. Since the darkonium mixes with the scalar mediator, existing searches for (displaced) $B \to K \phi,\ \phi \to \text{SM}$ decays of a Higgs-mixing scalar $\phi$ are sensitive to this decay topology. For a maximum branching ratio $\mathcal{B}(B \to K h_d) = 100\%$, the bound on the Higgs mixing angle from LHCb's di-muon search~\cite{LHCb:2016awg} can be interpreted as a bound on the combination
\begin{align}
    |\theta_h\theta_{h_d}| < 10^{-3} - 10^{-4}\qquad \text{for} \qquad m_{h_d} \in [250,4700]\,\text{MeV}\ .
\end{align}
This bound limits the overall rate of $h_d$ production through any channel. 

In general, however, $h_d$ decays to SM particles compete with decays into light binding-force carriers. Depending on the underlying model, searches for missing energy in two-body decays $B \to K \phi,\ \phi\to \slashed{E}$ are sensitive to decays into force carriers which remain invisible to the detector, see BaBar's search in Ref.~\cite{BaBar:2013npw} and the interpretation thereof in Ref.~\cite{Filimonova:2019tuy}. The recent evidence for $B \to K \nu\bar{\nu}$ in the Standard Model found by the Belle II collaboration~\cite{Belle-II:2023esi} shows that an analysis of the data in terms of two-body decays $B \to K \slashed{E}$ is within reach. Combining both SM and invisible final states maximizes the sensitivity to light scalar darkonia.

\paragraph{Production of stable darkonia}
The pseudo-scalar and vector darkonia $\eta_d$ and $\Upsilon_d$ can only be produced in pairs, due to the assumed discrete symmetries in the dark sector.

For the production from Higgs decays through $h \to \eta_d\eta_d$ and $h \to \Upsilon_d\Upsilon_d$, the decay rates are given by~\eqref{eq:hd-to-etad-etad} and~\eqref{eq:hd-to-upsd-upsd}, multiplied by $\theta_h^2$. They feature the same $|R_{h_d}'(0)|^4$ scaling as $h \to h_dh_d$ and are therefore of similar magnitude. Since $\eta_d$ and $\Upsilon_d$ are stable, they appear invisible to the LHC detectors and can in principle be probed in searches for invisible Higgs decays. However, due to the small branching ratio compared to $h\to SS$ and $h \to h_d S$, current searches~\cite{ATLAS:2023tkt} are not sensitive to $h \to \eta_d\eta_d$ and $h \to \Upsilon_d\Upsilon_d$ decays.

In top-antitop processes and $B\to K$ decays, the production of $\eta_d$ and $\Upsilon_d$ pairs is phase-space and/or mixing suppressed compared to $S$ and $h_d$. Searches for $t\bar{t}\slashed{E}_T$ or $B\to K\slashed{E}$ could probe the $S \eta_d \eta_d$ and $S \Upsilon_d \Upsilon_d$ couplings, but at a lower sensitivity than for invisible $h_d$ decays.

\subsection{$\mbox{F}_{\rm P}$ model}
As mentioned before, the $\mbox{F}_{\rm P}$ model requires a nontrivial condition for the mediator and dark fermion masses~\eqref{eq:FP-model-constraint} for the scenario where the binding force is also the mediator to the Standard Model. We will not restrict ourselves to this minimal scenario, but will consider a more general scenario with an additional force that binds the dark fermions into the $\eta_d$ bound state. This general scenario is characterized by the four parameters $m_P$, $m_{\eta_d}$, $\lambda_P$ and $\theta$. 

\paragraph{Production}
The production of $\eta_d$ and $P$ at the LHC mainly comes from Higgs boson decays. For $m_P, m_{\eta_d} < m_h$, the new decay channels for the SM Higgs boson are
\beqa
\Gamma(h \rightarrow P P ) &=& \frac{\lambda_P^2\,\cos^4{\theta}\,v^2}{2\pi\,m_h}\sqrt{1- \frac{4m_P^2}{m_h^2}} ~,  \quad
\Gamma(h \rightarrow \eta_d \eta_d ) = \frac{\lambda_P^2\,\sin^4{\theta}\,v^2}{2\pi\,m_h}\sqrt{1- \frac{4m_{\eta_d}^2}{m_h^2}} \nonumber \\
\Gamma(h \rightarrow P \eta_d ) &=& \frac{\lambda_P^2\,\sin^2{\theta}\cos^2{\theta}\,v^2}{\pi\,m_h}\,\frac{2}{m_h}|\vec{p}|(m_h, m_P, m_{h_d}) ~, \nonumber
\eeqa
with $|\vec{p}|$ from~\eqref{eq:momentum}.
In the limit of $m_P, m_{\eta_d} \ll m_h$, the sum of the three channels provides a new contribution to the total Higgs boson decay width, $\Gamma_h^{\rm NP} = \lambda_P^2 v^2/(2\pi m_h)$. Using the upper bound on the invisible Higgs branching ratio from~\eqref{eq:BRinv} for $\theta = 0$, $\text{BR}_{\rm inv} < 0.078$, we constrain the portal coupling~\footnote{This bound is comparable to that derived from direct searches for invisible Higgs decays~\cite{ATLAS:2023tkt}, $\text{BR}_{\rm inv} < 0.107$, yielding $\lambda_P < 0.0025$ at the 95\% C.L. The latter applies if both $\eta_d$ and $P$ are detector-stable.}
\begin{align}
\lambda_P < 0.0021 \qquad \text{at } 95\%\,\text{C.L.}
\label{eq:FP-invisible-Higgs-decay-constraint}
\end{align}

\paragraph{Decay}
Depending on the mass spectrum, the two parity-odd states $P$ and $\eta_d$ can decay into each other. In analogy with the $\mbox{F}_{\rm S}$ and $\mbox{F}_{\rm V}$ models, we choose the mass hierarchy $m_P < m_{\eta_d}$. In this case, the state $P$ can be a collider-stable particle, while $\eta_d$ decays via
\begin{align}\label{eq:fv-decays}
\eta_d & \to P\,h^{(\ast)} \to (P\,b\,\bar{b}), (P\,c\,\bar{c}), (P\,\tau^-\,\tau^+),  (P \, g\,g), (P\,\gamma\,\gamma) ~.
\end{align}
Here the specific channels depend on the relation of the mass splitting $\Delta m \equiv m_{\eta_d} - m_P$ and two-body final state mass sum from the off-shell Higgs boson decay. 

\begin{figure}[t!]
\centering
    \includegraphics[width=0.48\textwidth]{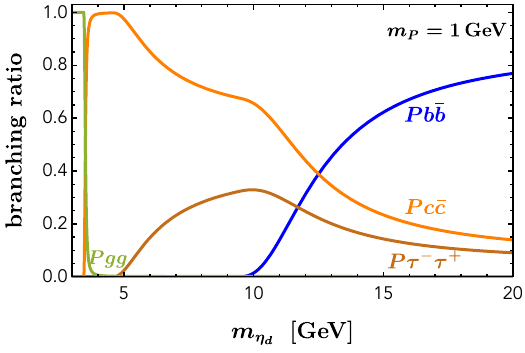} \hspace{3mm}
    \includegraphics[width=0.48\textwidth]{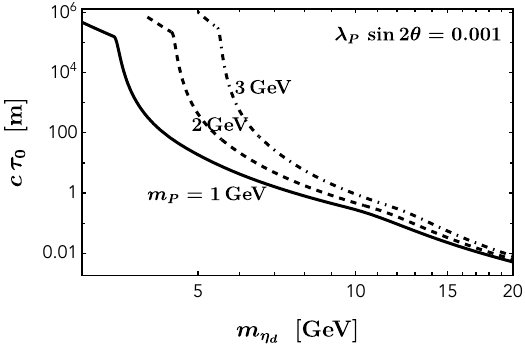}
    \caption{{\it Left panel:} The decay branching ratios of $\eta_d$ as a function of its mass. {\it Right panel:} The decay length of $\eta_d$ as a function of its mass for three different values of $m_P$. Note that the mass difference between $m_{\eta_d}$ and $m_P$ is chosen to be above 2 GeV, such that the parton-level decay width in \eqref{eq:FP-eta_d-decay-gluon} can be trusted.
    \label{fig:FP-decay}}
\end{figure}

For the mass splitting $\Delta m > 2 m_f$ with $m_f$ as a fermion mass, the three-body decay width is
\begin{align}\label{eq:FP-eta_d-decay-fermion}
\Gamma(\eta_d \rightarrow P \,f \,\bar{f}) = c_f\,\frac{\lambda_P^2\,\sin^2(2\theta)}{64 \pi^3 m_{\eta_d}^2}\int^{\Delta m^2}_{4m_f^2}\!\! dm_{12}^2 \,\frac{m_f^2\,m_{12}^2}{m_h^4}\,\left(1 - \frac{4m_f^2}{m_{12}^2}\right)^{3/2}\,|\vec{p}|(m_{\eta_d}, m_{12}, m_P) ~,
\end{align}
with $|\vec{p}|$ given in \eqref{eq:momentum} and $c_f = 3\,(1)$ for quarks (leptons). Here, we have used the approximation of $\Delta m \ll m_h$ and ignored the momentum in the Higgs propagator. 

The three-body decay width into the SM gauge bosons reads
\begin{align} \label{eq:FP-eta_d-decay-gluon}
\Gamma(\eta_d \rightarrow P \,g\,g) = \frac{\alpha_s^2\lambda_P^2\sin^2(2\theta)}{2^9\,3^3\,\pi^5}\frac{m_{\eta_d}^5}{m_h^4} \left[1 + 28 r^2(1 - r^4) - r^8 
\right]
\end{align}
with the mass ratio $r = m_P/m_{\eta_d}$. The decay width for $\eta_d \rightarrow P \gamma \gamma$ is smaller by the same ratio as the SM Higgs decay widths, $\Gamma(h\rightarrow \gamma\gamma)/\Gamma(h\rightarrow gg)\approx 0.028$.

In the left panel of Fig.~\ref{fig:FP-decay}, we show the branching ratios of $\eta_d$ as a function of $m_{\eta_d}$ while fixing $m_P = 1$~GeV. The fermion masses are set to $m_\tau = 1.78$~GeV, $m_c = 1.22$~GeV and $m_b = 4.19$~GeV~\cite{Peset:2018ria}. In the right panel, we show the decay length of $\eta_d$ as a function of mass for a fixed value of $\lambda_P \sin{2 \theta} = 0.001$ and different values of $m_P$. As one can see from this plot, $\eta_d$ could be a long-lived particle at colliders for $\lambda_P$ satisfying the invisible Higgs decay width constraint in \eqref{eq:FP-invisible-Higgs-decay-constraint}.

\begin{figure}[t!]
\centering
    \includegraphics[width=0.43\textwidth]{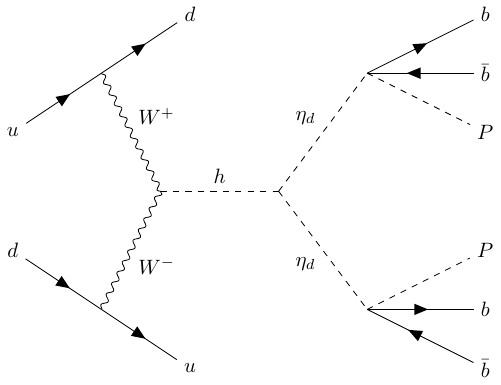} \hspace{10mm}
    \includegraphics[width=0.43\textwidth]{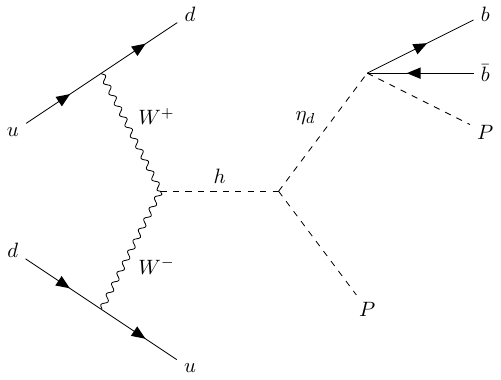}
    \caption{Feynman diagrams for the production of $2\eta_d$ and $\eta_d + P$ from Higgs boson decays produced via the vector-boson-fusion process at a hadron collider. The final particles $P$ are stable and appear as missing transverse energy $\slashed{E}_T$.
    \label{fig:FP-Feynman}}
\end{figure}

\paragraph{Signatures from Higgs decays}
Using Higgs production via vector boson fusion as an example, the novel signatures at the LHC are
\beqa
p p &\rightarrow& h jj \rightarrow \eta_d  \eta_d j j \rightarrow [P + b\bar{b}]_{\eta_d} [P + b\bar{b}]_{\eta_d} j j = 2(b\bar{b})
\,2j\,\slashed{E}_T  \label{eq:FP-signature-1} \\
p p &\rightarrow& h jj \rightarrow \eta_d  P j j \rightarrow [P + b\bar{b}]_{\eta_d} P j j = (b\bar{b})\,2j\,\slashed{E}_T ~, \label{eq:FP-signature-2}
\eeqa
with the corresponding Feynman diagrams shown in Fig.~\ref{fig:FP-Feynman}. 

For the benchmark scenario
\begin{align}
\label{eq:F_P-benchmark}
\lambda_P = 0.001\ ,\ \theta = \pi/4\ ,\ m_{\eta_d} = 15\,\text{GeV}\ ,\ m_P = 1\,\text{GeV}\ ,
\end{align}
the branching ratios of the new Higgs boson decay channels are
\beqa
\mathcal{B}(h \rightarrow P P ) = 0.47\%\,, \quad 
\mathcal{B}(h \rightarrow \eta_d \eta_d ) = 0.47\%\,, \quad 
\mathcal{B}(h \rightarrow P \eta_d ) = 0.94\%~.
\eeqa
The branching ratios of $\eta_d$ decays can be read from Fig.~\ref{fig:FP-decay} with $c\tau_0(\eta_d) = 0.023\,\mbox{m}$. Using again the 14-TeV VBF cross section of $4277\,$fb~\cite{LHCHiggsCrossSectionWorkingGroup:2016ypw}, we obtain the cross section for the signature~\eqref{eq:FP-signature-1} as $7.7\,$fb and for the signature~\eqref{eq:FP-signature-2} as $25\,$fb. In the first case, the signature consists of two displaced $b\bar{b}$ vertices and missing energy, while in the second case the signatures features one displaced $b\bar{b}$ vertex and missing energy. Notice that the displaced $b\bar{b}$ vertices do not reconstruct the mass of the bound state, since $\eta_d$ decays through the three-body process $\eta_d \to P + b\bar{b}$.

\paragraph{Signatures from $B$ meson decays} Since the pseudo-scalar mediator does not mix with the Higgs, darkonium production through Higgs-mixing as in the $\mbox{F}_{\rm S}$ model is not possible. Instead, darkonium or mediator pairs can be produced through the three-body decays $B \rightarrow K + h^* \rightarrow K + P\,P$, $B \rightarrow K + h^* \rightarrow K + \eta_d\,\eta_d$ and $B \rightarrow K + h^* \rightarrow K + P\,\eta_d$. For $B \rightarrow K + P\,P$, the signature is similar to $B \to K \nu \bar{\nu}$, with some differences in kinematics due to the finite mass of the stable $P$ particle in the final state. The mild excess reported by the Belle-II collaboration~\cite{Belle-II:2023esi} could be interpreted within the $\mbox{F}_{\rm P}$ model (see Ref.~\cite{Fridell:2023ssf,Bolton:2024egx} for a more general analysis of various possible interpretations). The interpretation of the $B \to K \nu \bar{\nu}$ excess in the $\mbox{F}_{\rm P}$ model involves correlated predictions for $B \rightarrow K + h^* \rightarrow K + \eta_d\,\eta_d$ and $B \rightarrow K + h^* \rightarrow K + P\,\eta_d$, as well the corresponding decay channels of the Higgs boson at the LHC, see \eqref{eq:FP-signature-1} and \eqref{eq:FP-signature-2}. In the minimal scenario, the final state of $\eta_d$ in these processes includes pairs of jets or leptons and missing energy, see~\eqref{eq:FP-eta_d-decay-fermion} and~\eqref{eq:FP-eta_d-decay-gluon}. In the general scenario, $\eta_d$ can be stable or decay into dark force carriers, leading to possibly different signatures.

\subsection{$\mbox{F}_{\rm V}$ model}\label{sec:fv-colliders}
The phenomenology of this model is determined by the dark-sector masses $m_{\eta_d}$, $m_{\Upsilon_d}$, $m_d$, as well as the mediator couplings to darkonia, $g_d$, and to SM particles, $\epsilon$. We assume the mass hierarchy $m_{\Upsilon_d} \gtrsim m_{\eta_d} > m_d$. The masses of the two bound states are almost degenerate. They are only separated by hyperfine splitting and by mixing-induced corrections to the $\Upsilon_d$ mass, see~\eqref{eq:-fv-mass-ev}. These (squared) mass corrections can be induced through $\Upsilon_d$ mixing with the dark photon, scaling as $\theta_{\Upsilon}^2 \sim m_d^4/m_{\Upsilon_d}^4$ for $m_d \ll m_{\Upsilon_d}$, and with the $Z$ boson, scaling as $\theta_Z^2 \sim \epsilon^2 t_w^2$. For a light dark photon and small kinetic mixing, these corrections are small and the approximation in~\eqref{eq:fv-mass-mixing} applies.

\paragraph{Decay}
Neither of the two bound states is stable. The decays of the darkonia are determined by the interactions described in Sec.~\ref{sec:fv-model}. The main decay modes of $\eta_d$ and $\Upsilon_d$ are
\begin{align}\label{eq:fv-decays-all}
\eta_d & \rightarrow \gamma_d\gamma_d ~, \\\nonumber
\Upsilon_d & \rightarrow \gamma_d\gamma_d\gamma_d\ , \quad
\Upsilon_d \rightarrow f \bar{f}\ , \quad 
\Upsilon_d \rightarrow \pi^0 \gamma\ , \quad 
\Upsilon_d \rightarrow \eta_d\gamma_d^{(\ast)}~.
\end{align}
In most of the parameter space, the $\Upsilon_d$ decays dominantly into three dark photons. For $m_{\Upsilon_d} \gtrsim 3 m_d$, i.e., close to the kinematic threshold, the decay is phase-space suppressed and other decay processes can be relevant.
The decay $\Upsilon_d \to \eta_d \gamma_d^{(\ast)}$ is always phase-space suppressed due to the small mass splitting between $\Upsilon_d$ and $\eta_d$. The dark photon in this decay can be off-shell, as indicated by an asterisk, leading to 3-body decays. The decay $\Upsilon_d \to \pi^0\gamma$ is induced through kinetic mixing, see~\eqref{eq:pion-couplings}. It typically has no phase-space suppression and can dominate over $\Upsilon_d \to \eta_d \gamma_d^{(\ast)}$, provided that the kinetic mixing is not too small.

The dark photon dominantly decays through kinetic mixing into lepton pairs or hadrons~\cite{Ilten:2018crw}. The anomaly- and mixing-induced decay $\gamma_d \rightarrow \pi^0 \gamma$ deduced from the BSEFT coupling in~\eqref{eq:pion-couplings} is relatively suppressed by $\alpha/(4\pi)^3$ and will be neglected.

\paragraph{Production}
The production of the darkonium states proceeds through the vector mediator. At $e^+e^-$ colliders, the lowest bound states can be produced through the processes (see also~\cite{An:2015pva})
\begin{align}
e^+ e^- \to \gamma_d^\ast \to \eta_d \gamma_d\ ,\qquad 
e^+ e^- \to \gamma_d^\ast \gamma \to \Upsilon_d\gamma\ .
\end{align}
In the production of $\eta_d$, the dark photon in the final state is radiated from the dark fermions before the bound state forms. Producing the vector bound state $\Upsilon_d$ requires the emission of a photon from the initial state. An exception are bound states with masses close to a $b\bar{b}$ resonance $\Upsilon(nS)$, which can be produced through mixing with the bottomonium at the $B$ factories.

At the LHC, possible production channels are (see also~\cite{Krovi:2018fdr})
\begin{align}
\text{Drell-Yan:} & \qquad pp \to \gamma_d^\ast \to \eta_d \gamma_d  & pp & \to \gamma_d^\ast \to \Upsilon_d \\\nonumber
\text{Higgs decays:} & \qquad h \to \eta_d\gamma_d Z & h & \to \Upsilon_d Z\\\nonumber
\text{Weak boson fusion:} & \qquad pp \to \eta_d \gamma_d jj & pp & \to \Upsilon_d jj\ .
\end{align}
Light darkonia can also be produced from bremsstrahlung through $q\to \gamma_d^\ast q \to \Upsilon_d q$. The production of an $\eta_d\gamma_d$ pair is phase-space suppressed.

\begin{figure}[t!]
\centering
    \includegraphics[width=0.48\textwidth]{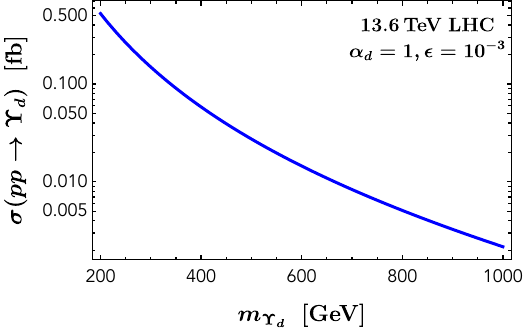} 
    \caption{Production cross section of $\Upsilon_d$ at the LHC via the Drell-Yan process $pp \rightarrow \Upsilon_d$.
    \label{fig:FV-cross-section}}
\end{figure}

All production cross sections are suppressed as $\epsilon^2$, because the bound states are produced through kinetic mixing. For $1 < m_d < 100\,$\,GeV, current bounds on kinetic mixing range around $\epsilon \lesssim 10^{-3} - 10^{-4}$~\cite{Batell:2022dpx}, which limits the expected event rates. In Fig.~\ref{fig:FV-cross-section} we show the cross section for $pp\to \Upsilon_d$, calculated by using the analytical formula for $Z^\prime$ production~\cite{Paz:2017tkr} and MSTW parton distribution functions~\cite{Martin:2009iq}. Moreover, using~\eqref{eq:eps-d} for the Coulomb case $m_d = 0$, we have determined the $\Upsilon_d$ coupling to fermions as $e\,\epsilon\,g_d\,\epsilon_d \approx e\,\alpha_d\,\epsilon/2$ in the limit $m_d \ll m_{\Upsilon_d}$.

\begin{figure}[t!]
\centering
    \includegraphics[width=0.5\textwidth]{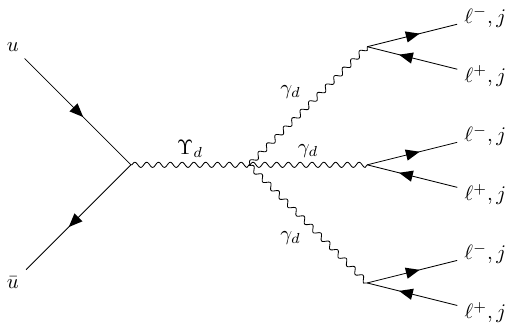} 
    \caption{Feynman diagram for the Drell-Yan production of $\Upsilon_d$ that cascade-decays into SM fermions via intermediate dark photons $\gamma_d$.
    \label{fig:FV-Feynman1}}
\end{figure}

\paragraph{Signatures at the LHC}
At the LHC, many different signatures with comparable event rates are possible. Drell-Yan production leads to the following signatures
\begin{align}\label{eq:fv-drell-yan}
pp & \rightarrow \eta_d \gamma_d \to [(\text{SM})_{\gamma_d}(\text{SM})_{\gamma_d}]_{\eta_d} (\text{SM})_{\gamma_d} \\\nonumber
pp & \rightarrow \Upsilon_d \to [(\text{SM})_{\gamma_d}(\text{SM})_{\gamma_d}(\text{SM})_{\gamma_d}]_{\Upsilon_d} 
\\\nonumber
pp & \rightarrow \Upsilon_d \rightarrow [\eta_d \gamma_d^\ast]_{\Upsilon_d} \to \big[[(\text{SM})_{\gamma_d}(\text{SM})_{\gamma_d}]_{\eta_d}\, \text{SM}\big]_{\Upsilon_d}\ ,
\end{align}
where $\text{SM} = \{\ell^+\ell^-,jj\}$. Depending on the mass, the intermediate dark photon can be on-shell or off-shell. In the case of a resonance, the decay products reconstruct the dark photon mass. For illustration, we show the Feynman diagram for the second process of \eqref{eq:fv-drell-yan} in Fig.~\ref{fig:FV-Feynman1}.
 The third process in~\eqref{eq:fv-drell-yan} can be relevant if the decays $\Upsilon_d \to 3\gamma_d$ and $\Upsilon_d \to f\bar{f}$ are relatively suppressed. Since $\Upsilon_d \to \eta_d\gamma_d^\ast$ is also kinematically suppressed by the small mass splitting between the two bound states, the decay products of $\Upsilon_d$ are likely displaced. Searches for (displaced) lepton pairs and/or multi-jets in association with missing energy, as well as mono-jet searches can be sensitive to such signatures.

For the benchmark scenario with 
\beqa
m_{\Upsilon_d} = 400~\mbox{GeV}\,,~ m_d = 20~\mbox{GeV}\,,~ \epsilon = 10^{-3} \,,~\alpha_d = 1 ~, 
\eeqa
one has the branching ratios
\beqa
&&\mathcal{B}(\Upsilon_d \rightarrow \gamma_d \gamma_d \gamma_d) \approx 100\% \\\nonumber
&& \mathcal{B}(\gamma_d \rightarrow \ell^+ \ell^-) \approx 30\%\,,\quad \mathcal{B}(\gamma_d \rightarrow \tau^+ \tau^-) \approx 15\%\,, \quad \mathcal{B}(\gamma_d \rightarrow jj) = 55\% ~. 
\eeqa
The production cross section at the LHC via the Drell-Yan process is $\sigma = 0.058$~fb. Including the branching ratio into jets, the rate for $pp \rightarrow \Upsilon_d \rightarrow \ell^+ \ell^- jjjj$ is about $\sigma\cdot\mathcal{B} \approx 0.016$~fb, which could be searched for at the HL-LHC with a luminosity of $3\,\mbox{ab}^{-1}$.

Weak boson fusion leads to similar signatures as Drell-Yan production. The two characteristic forward jets can be used for tagging.

Higgs decays offer an interesting alternative for searches, as long as the dark-sector states are sufficiently light. In this case, bound states can be produced through kinetic mixing of $\gamma_d$ or $\Upsilon_d$ with the $Z$ boson. Signatures are the exotic Higgs decays
\begin{align}
h & \rightarrow \eta_d \gamma_d Z \to [(\text{SM})_{\gamma_d}(\text{SM})_{\gamma_d}]_{\eta_d} (\text{SM})_{\gamma_d} (\text{SM})_Z\\\nonumber
h & \rightarrow \Upsilon_d Z \to [(\text{SM})_{\gamma_d}(\text{SM})_{\gamma_d}(\text{SM})_{\gamma_d}]_{\Upsilon_d} (\text{SM})_Z\\\nonumber
h & \rightarrow \Upsilon_d Z \rightarrow [\eta_d \gamma_d^\ast]_{\Upsilon_d} Z \to \big[[(\text{SM})_{\gamma_d}(\text{SM})_{\gamma_d}]_{\eta_d} \, \text{SM}\big]_{\Upsilon}\, (\text{SM})_Z\ .
\end{align}
The final-states particles of the first and third process are identical, but the kinematics is different. In the third process, the decay of $\Upsilon_d$ is suppressed by the small mass splitting, indicating displaced decays.

\paragraph{Signatures at Belle II} At Belle II, prominent signatures are similar to Drell-Yan production at the LHC, but with an associated photon instead of a jet:
\begin{align}
e^+ e^- & \to \eta_d \gamma_d \to [(\text{SM})_{\gamma_d}(\text{SM})_{\gamma_d}]_{\eta_d} (\text{SM})_{\gamma_d}\\\nonumber
e^+ e^- & \to \Upsilon_d\, \gamma \to [(\text{SM})_{\gamma_d}(\text{SM})_{\gamma_d}(\text{SM})_{\gamma_d}]_{\Upsilon_d}\,\gamma\\\nonumber
e^+ e^- & \to \Upsilon_d\, \gamma \to [\eta_d \gamma_d^\ast]_{\Upsilon_d}\, \gamma \to \big[[(\text{SM})_{\gamma_d}(\text{SM})_{\gamma_d}]_{\eta_d}\,\text{SM}\big]\, \gamma\ .
\end{align}
Again, depending on the lifetimes and masses of the involved particles, the signature can consist of (displaced) lepton pairs and/or multi-jets plus potential missing energy, or a mono-photon.

In all processes, the reconstruction of the bound states $\eta_d$ and $\Upsilon_d$ proceeds in the same way, provided that all decay products are detectable. In this case, the signatures only differ by the particles produced in association with the bound states. If final-state particles are missed because an intermediate state decays outside of the detector reach, signatures from different production channels and at different experiments can vary significantly, due to the different detector geometry and kinematics.

\section{Conclusions and outlook}
\label{SEC:conclusions}

In this work, we have analyzed the phenomenology of dark-sector bound states at colliders. To this end, we have developed the BSEFT framework, which allows us to calculate production and decay rates for on-shell darkonia for collider physics and beyond. The BSEFT factorizes amplitudes into high-energy contributions associated with the production and low-energy contributions associated with the formation of the bound state. For three minimal scenarios where the binding force carrier is identical with the portal mediator to the Standard Model, we have calculated the matching of the BSEFT coefficients to the underlying model. For other scenarios, this procedure can be repeated using the techniques outlined in this work.

Based on our BSEFT calculations, we have made predictions for darkonium signatures at colliders. The LHC and Belle II are complementary in probing darkonia with different masses and interactions with the Standard Model. In particular, darkonium production and decay lead to new signatures with multiple intermediate resonances in processes like Higgs boson and $B$ meson decays. Since the dark-sector interaction with the Standard Model is experimentally constrained, decays of the mediators are typically delayed, which leads to signatures with displaced lepton or jet pairs. In some models, stable darkonia add missing energy to the final state.

In our analysis, we have focused on the lowest-lying bound states in the darkonium spectrum, where the constituents are in s-wave or p-wave constellations. Extending the framework to excited states and states with higher angular momentum quantum numbers has interesting phenomenological consequences. Including excited states means producing an entire mass spectrum of darkonia at colliders, similar to the hadron spectra in the Standard Model. For darkonia with higher angular momentum, we expect new signatures at colliders, as both production and decay are sensitive to the CP quantum numbers of the bound states. As for s-wave and p-wave states, generalized parity and charge-conjugation symmetries in the dark sector can stabilize some of these states.

At future colliders, the scope of darkonium searches can be significantly extended. High-luminosity experiments like the FCC-ee~\cite{FCC:2018evy} or a muon collider~\cite{Accettura:2023ked} will allow to probe extremely rare signatures. For the scenarios considered in this work, these experiments will have an unprecedented sensitivity to darkonia with tiny couplings to the Standard Model. As a second benefit, future colliders will offer new search options for darkonia well above the weak scale~\cite{Bottaro:2021srh}. The FCC-ee or CEPC~\cite{CEPCStudyGroup:2018ghi} will extend the discovery potential of Belle II to much higher bound-state masses, providing an excellent environment to reconstruct signatures with displaced vertices and missing energy. At a muon collider, darkonia with scalar mediators can be produced through the new channels known for the SM Higgs boson, while darkonia with vector mediators benefit from efficient new production processes like vector boson fusion.

Collider searches for darkonia do not only offer new opportunities for a discovery. Interpreted in terms of dark matter scenarios, they can also help to reveal the fundamental nature of dark matter, independently of whether a signal is observed or not. A parti\-cularly interesting aspect is the correlation between bound-state formation in darkonium production at colliders and in dark matter annihilation in the early universe. By exploiting such correlations, collider searches can probe the cosmic history of dark matter in freeze-out scenarios. Another connection arises from the naturally light mediators in bound-state scenarios, which induce self-interactions among dark matter particles. Here, collider searches for darkonia can probe the impact of self-interactions on small-scale structures in galaxies and clusters, where observations differ from current prediction. We look forward to exploring these new connections between colliders and the cosmos.

\acknowledgments
We thank Yue Zhang for sharing his calculation of $\Upsilon_d \to 3\gamma_d$ with us. This work was initiated at the Aspen Center for Physics, which is supported by the National Science Foundation grant PHY-2210452. The work of YB is supported by the U.S. Department of Energy under the contract DE-SC-0017647.

\appendix
\section{BSEFT couplings and wave functions}
\label{APPEND:wave-function}

In this appendix, we derive the relations between some BSEFT couplings and the wave functions of the bound states. Following the notation in Ref.~\cite{Bodwin:1994jh}, we use the gamma-matrix convention in Bjorken and Drell~\cite{Bjorken:1965sts} and the Foldy-Wouthuysen-Tani (FWT) transformation~\cite{Foldy:1949wa,Tani:1951trl} to perform the non-relativistic expansion of the field operators. In this notation, one has $\alpha_i = \{\{0, \sigma_i\},\{\sigma_i, 0\}\}$ with $\sigma_i$ as the Pauli matrices and $\beta = \{\{\mathbb{I}_2, 0\},\{0, -\mathbb{I}_2\}\}$. The gamma matrices are $\gamma^0 = \beta$, $\gamma^i = \beta \, \alpha_i$ and $\gamma^5 = i\,\gamma^0\gamma^1\gamma^2\gamma^3$.  For a Dirac fermion at rest, the 4-component spinor is defined as $\Psi^T = (\psi^T, \chi^T)$ with the 2-component Pauli spinor $\psi$ to annihilate a heavy quark and $\chi$ the Pauli spinor to create a heavy antiquark. 

Under the FWT transformation, the Dirac spinor and the Hamiltonian transform as
\begin{align}
    \Psi' = e^{i\,S}\,\Psi~,\qquad H' = e^{i\,S} H e^{-i\,S},
\end{align}
with 
\beqa
e^{i\,S}  = \cos{(|\bm{p}|\,\theta)} + \frac{\beta\,{\bf \alpha}\cdot\,{\bm{p}}}{|{\bm{p}}|} \, \sin{(|\bm{p}|\,\theta)} ~,
\eeqa
where $\tan{(2\,|\bm{p}|\,\theta)} = |\bm{p}|/m$ and $m$ as the fermion mass.

For the $\mbox{F}_{\rm S}$ model, we are interested in the scalar bilinear of a fermion and an antifermion with the operator form $\mathcal{O}_{\rm S} \equiv \Psi^\dagger \gamma^0 \Psi = \Psi^\dagger \beta \Psi$. After the FWT transformation and performing an expansion in $|\bm{p}|/m$, the scalar operator contains 
\beqa
\mathcal{O}_{\rm S} \supset \frac{1}{m}\, \chi^\dagger \left(\tfrac{i}{2}\overset{\text{\tiny$\leftrightarrow$}}{\bm{\partial}}\cdot \bm{\sigma}\right) \psi \,+\, \frac{1}{m}\, \psi^\dagger \left(\tfrac{i}{2}\overset{\text{\tiny$\leftrightarrow$}}{\bm{\partial}}\cdot \bm{\sigma}\right) \chi ~, 
\eeqa
where $\chi^\dagger \overset{\text{\tiny$\leftrightarrow$}}{\bm{\partial}} \psi \equiv \chi^\dagger (\bm{\partial} \psi) - (\bm{\partial} \chi^\dagger) \psi$. Using a relativistic normalization for the bound states as in Ref.~\cite{Braaten:1996ix}, the matrix element between the scalar bound state $h_d$ and the vacuum reads~\footnote{See (3.19b) in Ref.~\cite{Bodwin:1994jh} for a non-relativistic normalization.}
\beqa
\langle 0 | \chi^\dagger \left(-\tfrac{i}{2}\overset{\text{\tiny$\leftrightarrow$}}{\bm{\partial}}\cdot \bm{\sigma}\right) \psi | h_d \rangle \approx \sqrt{2m_{h_d}}\frac{3}{\sqrt{2\pi}} \, R'_{h_d}(0) ~,
\eeqa
neglecting higher-order terms in velocity expansion.

Matching the $h_d - S$ mixing coefficient in the BSEFT Lagrangian~\eqref{eq:bseft-FS}, one has
\beqa
g_d\,\mu_d^2 = g_d \,\frac{\sqrt{2m_{h_d}}}{m_\chi}\, \frac{3}{\sqrt{2\pi}}\,R'_{h_d}(0) ~. 
\eeqa

For the $\mbox{F}_{\rm P}$ model and following the similar procedure, one has
\beqa
\mathcal{O}_{\rm P} = i \Psi^\dagger \gamma^0 \gamma^5 \Psi \supset -i\,(\chi^\dagger \psi - \psi^\dagger \chi) ~.
\eeqa
By identifying the matrix element of the pseudo-scalar bound state $\eta_d$ as
\beqa
\langle 0 | -i\,\chi^\dagger \psi | \eta_d\rangle \approx \sqrt{2m_{\eta_d}}\,\frac{1}{\sqrt{2\pi}}\,R_{\eta_d}(0)~, 
\eeqa
the $\eta_d - P$ mixing coefficient in the BSEFT Lagrangian~\eqref{eq:bseft-F_P} is matched to be 
\beqa
g_{d5}\,\mu_d^2 = 2\,g_{d5}\, \sqrt{\frac{m_{\eta_d}}{\pi}}\,R_{\eta_d}(0)~. 
\eeqa
%

\providecommand{\href}[2]{#2}\begingroup\raggedright\endgroup


\end{document}